\documentstyle[emulateapj]{article}
\begin{document}

\slugcomment{To appear in the September 2000 Astrophysical Journal Supplement Series}

\title{A Uniform Analysis of the Ly-$\alpha$ forest at $z=0 - 5$: \\ 
II. Measuring the mean intensity of the
extragalactic ionizing background using the proximity effect}

\author{Jennifer Scott\altaffilmark{1}, Jill Bechtold\altaffilmark{1}, 
Adam Dobrzycki\altaffilmark{2}, \& Varsha P. Kulkarni\altaffilmark{3}}

\altaffiltext{1}{Steward Observatory, University of Arizona, Tucson,
AZ 85721, USA,\\
e-mail: [jscott,jbechtold]@as.arizona.edu}
   
\altaffiltext{2}{Harvard-Smithsonian Center for Astrophysics,
60 Garden Street, Cambridge, MA 02138, USA,\\ e-mail:
adobrzycki@cfa.harvard.edu}

\altaffiltext{3}{Clemson University, Department of Physics and Astronomy,
Clemson, SC  29634, USA,\\ e-mail:
varsha@clemson.edu}

\begin{abstract}

Moderate resolution data for 40 quasi-stellar objects (QSOs) at z $\approx$ 2 
were combined with
spectra of comparable resolution of 59 QSOs with redshifts
greater than $1.7$ found in the literature to form a large, homogeneous sample
of moderate resolution ($\sim$1 $\AA$) QSO spectra.  These spectra
were presented and the statistics of the Lyman $\alpha$
forest were discussed in Paper I.
In this analysis,
we demonstrate that a proximity effect is present in the data, ie.
there exists a significant (5.5$\sigma$) deficit of lines at $z_{abs} \approx
z_{em}$.  Within 1.5 $h^{-1}$ Mpc of the QSO emission redshift, 
the significance does 
depend on QSO luminosity, 
in accordance with the theory that 
this effect is caused by enhanced ionization of hydrogen in the
vicinity of the QSO from UV photons from the QSO itself. 
The photoionization model of Bajtlik, Duncan, \& Ostriker 
permits an estimate of the mean intensity of the extragalactic
background radiation at the Lyman limit.   We compare the results of this
standard analysis with those obtained using a maximum likelihood technique.
If the spectrum of the background
is assumed to be identical to that of each individual QSO, and if 
this background is
assumed to be constant over the redshift range $1.7 < z < 3.8$, 
then the best fit value for $J(\nu_{0})$ is found to be
1.4$^{+1.1}_{-0.5}$ $\times$  10$^{-21}$ ergs s$^{-1}$ cm$^{-2}$ Hz$^{-1}$ sr$^{-1}$,
using QSO redshifts based on the Ly$\alpha$ emission line.
Systemic QSO redshifts based on the  [OIII] $\lambda$5007 
emission line for 19 objects in our sample
show an average redshift of $\sim$400 km s$^{-1}$ 
with respect to Ly$\alpha$ emission.
Using redshifts based on [OIII]  or Mg II for the 35 
objects for which they are measured and adding 400 km s$^{-1}$ to the
remaining QSO Ly$\alpha$ redshifts  gives a lower value of $J(\nu_{0})$, 
7.0$^{+3.4}_{-4.4}$ $\times$ 10$^{-22}$ ergs s$^{-1}$ cm$^{-2}$ Hz$^{-1}$ sr$^{-1}$.
This value is in reasonable agreement with the predictions of 
various models of the
ionizing background based on the integrated QSO luminosity function.
Allowing for the fact that individual QSOs have different spectral
indicies which may also be different from that of the background, we
use the standard methods to solve for the HI photoionization rate,
$\Gamma$, and the parameters describing its evolution with redshift.
The best fit value for the HI ionization rate we derive is 
1.9$^{+1.2}_{-1.0}$ $\times$ 10$^{-12}$ s$^{-1}$,
in good agreement with models of the
background which incorporate QSOs only. 
Finally,  we use
simulated Lyman $\alpha$ forest spectra including the proximity effect
to investigate curve-of-growth effects in the photoionization model
used in the analysis.  We find that the presence of  
lines on the saturated part of the curve-of-growth could cause our
estimates of the background intensity to be 
overestimated by a factor of two to three.
This large absorption line sample and these
techniques for measuring the background and understanding the systematics
involved allow us to place what we believe are the firmest limits 
on the background at these redshifts.

\end{abstract}


\section{Introduction}
The spectra of  quasi-stellar objects (QSOs) 
blueward of Ly$\alpha$ emission show a prodigious number
of absorption lines primarily due to Ly$\alpha$ absorption by intervening 
neutral hydrogen along the line of sight to the QSO (Lynds, 1971; Sargent
et al. 1980; Weymann, Carswell, \& Smith, 1981).
In models of these Ly$\alpha$ systems, they are in photoionization equilibrium
with a background ultraviolet radiation field.  
This radiation field is modeled  as
the integrated emission from QSOs and young galaxies (Bechtold et al. 1987;
Miralda-Escud\'{e} \& Ostriker 1990; Madau 1992, Meiksin \& Madau
1993, Madau \& Shull 1996, Haardt \& Madau 1996, Fardal et al. 1998).

The basic trend in the number density evolution of Ly$\alpha$ absorbers
(12 $\lesssim$ log(N$_{HI}$) $\lesssim$ 16 cm$^{-2}$) 
at high redshift in a given
QSO spectrum is a steep power law increase with redshift (Sargent et al. 1980).
However, a decrease in the
number of lines near the emission redshift of an individual QSO 
relative to the number of lines expected from the power law distribution 
has been observed (Carswell et al. 1982,
Murdoch et al. 1986; Tytler 1987;
Bajtlik, Duncan,
\& Ostriker 1988, hereafter BDO; Lu, Wolfe, \& Turnshek 1991, hereafter LWT;
Bechtold 1994, hereafter B94).  The simplest explanation for 
this $``$inverse" effect is enhanced ionization of HI in the
vicinity of the QSO by ultraviolet photons from the QSO itself.  Thus, 
the name $``$proximity effect" is also used.  This interpretation,
along with the assumptions about the spectrum of the background and 
the photoionization of the nearby intergalactic medium (IGM)
by the QSOs, allows 
for a measurement of the mean intensity of the ionizing background at the
Lyman limit of hydrogen (BDO), which can be compared to estimates of 
the integrated emission from QSOs.

BDO find that $J(\nu_{0})$ is approximately constant over a redshift range
$1.7 < z < 3.8$. Expressing $J(\nu_{0})$ as $J_{-21}$ $\times$ 10$^{-21}$ ergs
s$^{-1}$ cm$^{-2}$ Hz$^{-1}$ sr$^{-1}$, these authors find $J_{-21} \approx 1$.
Their best fit form for the dependence of $J(\nu_{0})$ on redshift rules 
out a luminosity dependent cutoff
in the QSO luminosity function (BWLM; Schmidt, Schneider,
\& Gunn 1986).
Several other authors have carried out this analysis on other data sets
(LWT, Kulkarni \& Fall 1993, hereafter KF93, B94, Williger et al. 1994, 
Cristiani et al, 1995, Giallongo et al. 1996.)
LWT found $J_{-21} \approx 1$ for $1.7 < z < 3.8$; and  B94 found a
value consistent with this, $J_{-21} \approx 3$ for $1.6 < z < 4.1$. 
Srianand \& Khare (1996) compile a sample of 69 QSOs from the literature
(54 from B94) and obtain $J_{-21} \approx 6$ for $1.7 < z < 4.1$.
Williger et al. (1994) find a lower value, $J_{-21} \approx 0.1-0.3$
for z $\sim$4.2 from a single z=4.5 QSO; and Giallongo et al. (1996) find 
$J_{-21} \approx 0.5$ for $2.8 < z < 4.1$, based on high resolution spectra
of six QSOs.
KF93 re-express the BDO formalism using maximum
likelihood  techniques and use this to derive $J(\nu_{0}) \approx$ 6 $\times$ 10$^{-24}$
ergs s$^{-1}$ cm$^{-2}$ Hz$^{-1}$ sr$^{-1}$
for z $\approx$ 0.5 from the sample of Bahcall et al. (1993).

In this paper, a homogeneous sample of
QSO spectra is used 
to measure the mean intensity of the ionizing
background via the standard proximity effect analysis and the
maximum likelihood analysis of KF93.
We have presented new Lyman $\alpha$ forest data for 39 objects with 
$1.9 < z_{em} < 2.5$ in Paper I; and supplemented
this sample with 59 objects from the literature, and one QSO from the
Hamburg/CfA Bright Quasar Survey (Dobrzycki, Engels, \& Hagen 1999).
The spectra comprising our dataset are of moderate resolution, $\sim$1 $\AA$ FWHM;
but this is not a disadvantage, as this analysis requires good absorption line statistics 
and therefore many QSO sight lines.  This is 
difficult to achieve at high resolution, the major reason for using a large 
set of moderate resolution spectra such as this one.

A major uncertainty in the proximity effect analysis is in the  
systemic redshifts of the QSOs. 
Redshifts measured from
low ionization permitted lines (e.g. Balmer lines or Mg II) or forbidden lines
(e.g. [OIII] $\lambda\lambda$4959,5007)
lines have been shown to be redshifted with respect to Ly$\alpha$ and C IV
emission by up to $\approx$250 km s$^{-1}$ (Boroson \& Green 1992, Laor et al.
1995).   B94 found that increasing the values of the QSO redshifts by
1000 km s$^{-1}$ caused the best fit value of $J_{-21}$ to be decreased
by a factor of 3.
We therefore obtained emission line spectra for several objects 
in our sample in order to examine
redshift differences between Ly$\alpha$ and [OIII] $\lambda\lambda$4959,5007,
Mg II or Balmer emission. 
We  investigate the effect of these shifts 
on the value of $J(\nu_{0})$ derived.

In Section~\ref{sec-data} the spectra used
to measure Lyman limit fluxes for some of our sample objects and spectra used
to perform
systemic redshift measurements of several
QSOs in our sample are presented.
In Section~\ref{sec-pe}
the techniques used to measure the mean intensity of the HI ionizing 
background at the Lyman limit are discussed, which includes discussions 
of the QSO systemic redshifts in Section~\ref{sec-zs} and the HI ionization
rate measurements in Section~\ref{sec-gamma}.
In Section~\ref{sec-cog} we present Lyman $\alpha$ forest simulations
and investigate curve-of-growth effects in the proximity effect analysis.
Section~\ref{sec-disn} is a summary and discussion of our results and 
the possible systematic effects entering into our analysis.

Throughout this paper, we assume values of 75 km s$^{-1}$ Mpc$^{-1}$ for
H$_{0}$ and 0.5 for $q_{0}$.

\section{Data}
\label{sec-data}

\subsection{Spectrophotometry}
Spectrophotometry of 12 sample objects in the spectral region
between Ly$\alpha$ and C IV emission was obtained at the Steward Observatory (SO)
Bok Telescope with the Boller and Chivens (B\&C) Spectrograph 
and the 1200 $\times$ 800 CCD on the nights
of September 22, 1992, November 29, 1994, and March 28, 1995.  
Observations were made with a 400 l mm$^{-1}$ grating with $\lambda_{b}$=4889
$\AA$ in the first order and a 4.5$\arcsec$ slit.
Spectrophotometry of the object 1422+231 was obtained 
at the SO B\&C using a  600 l mm$^{-1}$ grating with
$\lambda_{b}$=6681 $\AA$ in the first order and a 
1.5$\arcsec$ slit on April 22, 1996; and
the object 1603+383 was observed by A.D. as part of the
Hamburg/CfA Bright Quasar Survey on July 4, 1995
with the Fred Lawrence Whipple Observatory 1.5-meter Tillinghast telescope
and FAST
spectrograph, using a 300 l mm$^{-1}$ grating with $\lambda_{b}$=4750
in the first order and a 3$\arcsec$ slit.
See Table~\ref{table-specphotobs} for a summary.

All observations except those of
1422+231 and 
1603+383 were made with the
slit set at the parallactic angle.  This should not seriously 
effect the spectrophotometry
of 1603+383 as it was observed at a small airmass.
Additionally, however, the observation
of 1422+231 was made with a slit width that is somewhat small for optimal 
spectrophotometry.  
In any case, as discussed further below, both 1422+231 and 1603+383 are 
excluded from the
proximity effect analysis due to the fact that 1422+231 is a gravitational
lens and the presence of associated absorption in the spectrum of 
1603+383.
Any small errors in the spectrophotometry of the 74 objects used in
the proximity effect analysis should not
significantly bias the results of this work.

\begin{figure*}
\epsscale{1.00}
\plotone{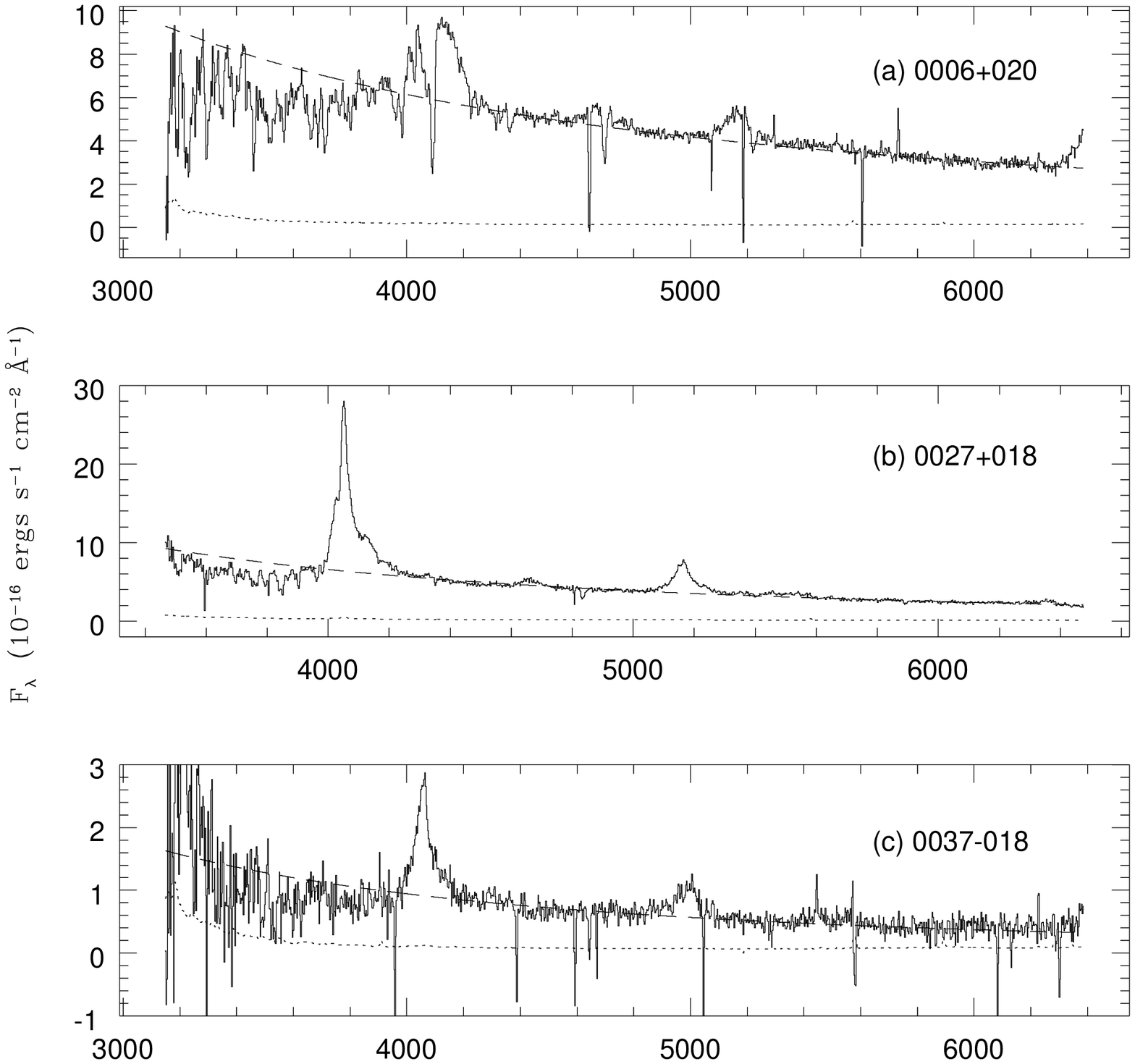}
\caption
{Spectrophotometry of z $\approx$ 2 QSOs; Dashed line indicates
the power law continuum fit; dotted line indicates the 1$\sigma$ errors
\label{fig:specphot} }

\end{figure*}

Object spectra were bias corrected and extracted using standard IRAF 
packages using He-Ne-Ar
and quartz calibration exposures taken at each telescope position to
perform  the wavelength calibration and to correct for pixel-to-pixel 
variations, respectively.  The data were then flux calibrated
using standard star exposures.  The column density of Galactic neutral 
hydrogen  along the line of sight to each object was found using the 
program COLDEN, made available by J. M$^{\rm c}$Dowell; and the spectra
were corrected for the Galactic reddening  calculated from the
relation $N_{HI}/E(B-V)$ = 4.8 $\times$ 10$^{21}$ atoms cm$^{-2}$ magnitude$^{-1}$
(Bohlin et al. 1978).  The spectra and the power law continuum fits are shown
in Figure~\ref{fig:specphot}.

\subsection{QSO Systemic Redshifts}
For the present absorption line sample, 
the QSO narrow emission lines discussed above
all lie redward of $\sim$7600 $\AA$, and into the near
infrared.
Spectra of four objects in this sample
were obtained at the MMT with the infrared spectrometer FSpec
(Williams et al. 1993) on May 20, 1994   (1207+399 and 1422+231) and
April 1, 1996 (1408+009, and 1435+638) using a 75 l mm$^{-1}$ grating
and a 1.2$\arcsec$ slit giving a resolution of $\sim$34 $\AA$ in the K
band.  A series of exposures of each object was taken.  Between each
exposure,
the object was moved along the slit.  The total integration time is
listed in Table~\ref{table-z2obs}.
One object, 0836+710, was observed on March 28, 1995
with the B\&C, the 1200x800 CCD, a 300
l mm$^{-1}$ grating with $\lambda_{b}$=6693 $\AA$ in the first order,
and a 4.5$\arcsec$ slit.
Infrared spectra of eight objects in this sample, 
0000-263, 0014+813, 
0636+680, 0956+122, 1159+124, 1208+101, 2126-158, were obtained using
FSpec, OSIRIS on the Cerro Tololo Inter-American Observatory
4 m telescope, and CRSP on the Kitt Peak National Observatory 4 m telescope
(Kuhn 1996).
A summary of these observations is given in Table~\ref{table-z2obs} and
the spectra are displayed in Figure~\ref{fig:nlz}.

\section{Ly$\alpha$ Forest Statistics for $z_{abs} \approx z_{em}$: The
	 Proximity Effect}
\label{sec-pe}

\subsection{Spectrophotometry}
\label{sec-pespecphot}
In order to perform the proximity effect analysis, the flux
of each QSO at the Lyman limit is needed.  The spectrophotometry data
discussed above  was used for this purpose.
A power law of the form $f_{\nu} \sim \nu^{-\alpha}$ was fit to the continua
of these objects.  The straight line fit to log($f_{\nu}$) vs. log($\nu$)
was done using a robust estimation technique; and
emission lines found by visually inspecting the spectrum  
were excluded from the points used in the fit.
The measured flux at 1450 $\AA$ and the value of $\alpha$
derived from this fit were used to determine the flux at 912 $\AA$.
For the objects we did not observe,
we proceed as follows.  If a flux measurement
at a rest UV wavelength other than 912 $\AA$ exists along with a published
spectral index, we use these to extrapolate to the Lyman limit.  If no
spectral index is available, we use the value of $0.46$ (Francis 1996).
The object 2134+004 has a variable continuum (Perez et al. 1989, Corbin 1992). 
Therefore, 
although we have spectrophotometry from our own observations of this object, 
we take the flux measurement of these authors from their averaged spectrum
produced from observations made over several months.  We use this with the
spectral index we derive to extrapolate to 912 $\AA$. 

If no rest UV spectrophotometry of an object exists, we estimate $f_{\nu}$
at 5500 $\AA$ (observed) from the V magnitude
given in Table 1 of Paper I with an extinction correction applied.
The extinction correction was calculated using the 
column density of neutral hydrogen 
from COLDEN and the Seaton (1979) re-normalization of the
composite UV-optical reddening curve of Nandy et al. (1975, and references
therein).
A rest-frame composite QSO spectrum (Zheng et al. 1997) with
an arbitrary flux scale was redshifted
by the appropriate amount for each object.  The flux in the V filter was
calculated by convolving this spectrum with the V filter transmission
as a function of wavelength.  A scaling factor was calculated so that
when the redshifted QSO composite spectrum was multiplied by this factor,
the resulting magnitude matched the magnitude listed in 
Table 1 of Paper I.  The flux at 1450 $\AA$ was then taken from
this scaled spectrum and this flux was extrapolated to the Lyman limit
using the spectral index given in Table~\ref{table-specphotfinal}.
A zero point flux density for
the V filter of 3.81 $\times$ 10$^{-20}$ ergs s$^{-1}$ cm$^{-2}$ Hz$^{-1}$
(Johnson 1966) was used. 

The asterisks in Table~\ref{table-specphotfinal} mark 
QSOs which are known lenses or which show associated absorption in
their spectra.  Associated absorption is 
defined to be any Lyman $\alpha$ absorption within
$\sim$5000 km s$^{-1}$ of the QSO redshift which also shows metal lines.
(See Paper I for a description of the metal line systems identified in each
QSO spectrum.)
These objects were excluded from the proximity effect analysis on the grounds
that gas associated with the QSO or QSO host galaxy is not part of 
the general 
intergalactic medium and bulk motions within this gas may skew the results.
The spectrophotometric properties adopted for the 59 QSOs from the
literature are listed in Table 5 of B94.

\subsection{Number of Lines with $z_{abs} \approx z_{em}$}
\label{sec-npr}

The first method we use to demonstrate the proximity effect is to
compare the number of lines predicted if there was no effect from 
the equation
\begin{equation}
\frac{d{\cal N}}{dz}= {\cal A}_{0}(1+z)^{\gamma} \label{eq:dndz} .
\end{equation}
with the number of lines counted in the spectrum as
a function of distance from the QSO,
\begin{equation}
\Delta {\cal N} = {\cal N}_{pred}-{\cal N}_{obs}. 
\end{equation}
The number of lines predicted is found by integrating Equ.~\ref{eq:dndz},
\begin{equation}
{\cal N}_{pred} = \frac{{\cal A}_{0}}{\gamma + 1} ((1+z_{max})^{\gamma+1} -
(1+z_{min})^{\gamma+1}) \label{eq:np}. 
\end{equation}
The bins in luminosity distance from the QSO are defined according to the 
relation,
\begin{equation}
\Delta R = 1687.5 \frac{\Delta z}{(1+z_{em})^{5/2}}  h^{-1} \; \; {\rm Mpc}.
\end{equation}
We use $h=0.75$.
Figure~\ref{fig:zlum} plots 
the distribution in z and Lyman limit luminosity of the
QSOs in our sample.

The dataset was divided into low luminosity and high luminosity subsamples
at log[$L(\nu_{0})$]=31.1,
such that there were equal numbers of objects in each subsample.  The
Lyman limit luminosity of each object was calculated according
to the expression
\begin{equation}
L(\nu_{0}) = 4\pi d_{L}^{2} \frac{f(\nu_{0})}{(1+z_{em})} 
\end{equation}
where the luminosity distance to the QSO, $d_{L}$ is given by
\begin{equation}
d_{L}=\frac{c \{q_{0}z + (q_{0}-1)[(1+2q_{0}z)^{1/2}-1]\}}{q_{0}^{2}H_{0}}
\end{equation}
for $q_{0}>0$.  In this paper, we use a value of 0.5 for $q_{0}$.
Figure~\ref{fig:npred} plots the fractional deficit of lines, 
\( ({\cal N}_{pred}-{\cal N}_{obs})/{\cal N}_{pred}\),
for the total sample and the high and low luminosity subsamples.

For the total sample, a 5.5$\sigma$ deficit
of lines is found in the 0-1.5 $h^{-1}$ Mpc bin.  The low luminosity subsample
shows a deficit of lower significance (3.6$\sigma$) than the high
luminosity subsample (4.6$\sigma$).  
These deficits are expected for 
a proximity effect caused by enhanced
ionization of HI from the QSO flux; and the marginally higher 
significance for high luminosity objects further suggests that this
picture is legitimate.

\subsection{Photoionization Model}
We follow the formalism outlined in BDO to calculate a value of the
mean intensity of the ionizing background in the redshift range
$1.7 < z < 3.4$.
The column density of a Ly$\alpha$ absorber in the immediate vicinity of a 
QSO will  be modified from the value that it would have if the QSO 
were not present.  The amount by which the column density of HI will be 
reduced due to ionization by UV photons from the QSO is given by
\begin{equation}
N=N_{0}(1+\omega)^{-1} \label{eq:column}
\end{equation}
where $N$ is the observed column density of the absorber, and $N_{0}$ is
the column density that the absorber would have if the QSO were absent.
The column density distribution of the general Ly$\alpha$ absorber
population was been shown to follow a power law over several orders
of magnitude in column density,
\begin{equation}
{\cal N} \propto  N^{-\beta},
\end{equation}
which, for a fixed limiting column density, $N_{thr}$, (corresponding
to the limiting rest equivalent width) can be integrated to give the
total number of lines with column densities equal to or larger than
the limiting value, \({\cal N}(N \geq N_{thr}) = N_{thr}^{-(\beta - 1)} \).
Thus, a proximity effect-corrected redshift distribution for a fixed
rest equivalent width threshold can be derived:
\begin{equation}
\frac{d{\cal N}}{dz}= {\cal A}_{0}(1+z)^{\gamma}[1+\omega(z)]^{-(\beta-1)}
\end{equation}
where $\omega$ represents a flux-scaled distance of each cloud from the 
QSO
\begin{equation}
\omega = \frac{F^{Q}(\nu_{0})}{4\pi J(\nu_{0})}. \label{eq:omega}
\end{equation}
Here, \(F^{Q}(\nu_{0})\) is the Lyman limit flux density due to the QSO
at the position of a given absorber, 
\begin{equation}
F^{Q}(\nu_{0})= \frac{L(\nu_{0})}{4\pi r_{L}^{2}}
\end{equation}
where $r_{L}$ is the luminosity distance between the QSO and 
the absorber.
We remove the dominant dependence
of the line density on redshift by introducing a coevolving coordinate,
$X_{\gamma}$, given by
\begin{equation}
X_{\gamma} = \int (1+z)^{\gamma} dz. 
\end{equation}
If no proximity effect existed, the number of lines per coevolving coordinate
would be expressed as
\begin{equation}
d{\cal N}/dX_{\gamma} = {\cal A}_{0} \label{eq:nope} .
\end{equation}

In this analysis, we use a value for $\beta$ of 1.46 from
of Hu et al. (1995) based upon high S/N, high resolution  spectra of
four QSOs at z $\approx$ 3, consistent with the value of 1.4 found
by  Dobrzycki \& Bechtold (1996), hereafter DB96, 
from simulations of Ly$\alpha$ forest
spectra in QSOs at z $\approx$ 3.  The value of this parameter is an
important factor in the ionization model.  B94 found that changing the
adopted value of $\beta$ from 1.7 to 1.4 caused the derived value of
$J_{-21}$ to decrease by a factor of $\sim$3.  Giallongo et al. (1996) find
that a double power law provides a better fit to the observed column density
distribution in their high resolution spectra than a single power law.
The form of their double power law consists of a break at 
$N_{HI}=10^{14}$ cm$^{-2}$ and values of $\beta$ above and below this
break of 1.8 and 1.4 respectively.  For this analysis, however, we will
use a single power law, as the data of Hu et al. (1995) do not require
the double power law form.

The procedure consists of assuming a form for $J(\nu_{0})$ as a function
of z, 
dividing the lines into the appropriate $\omega$ bins, and finding the
parameters of the assumed form of $J(\nu_{0})$ that gives the lowest 
$\chi^{2}$ between the binned data and the ionization model.
Since no work to date has shown that $J(\nu_{0})$ evolves significantly with
redshift over the range of our sample objects, 
we will treat the case that $J(\nu_{0})$  is
constant over the redshift range of the data.  

Figure~\ref{fig:chi2} plots $\chi^{2}$ with respect to the constant $J(\nu_{0})$
photoionization model versus log[$J(\nu_{0})$] and Figure~\ref{fig:dndx} plots
the coevolving number density versus $\omega$ for the lowest
$\chi^{2}$ value of $J(\nu_{0})$ for each subsample. 
The results of this analysis are summarized in Table~\ref{table-jnu} and are
discussed in more detail in Section~\ref{sec-disn}.

\subsection{Maximum Likelihood Analysis}

In addition to the standard BDO analysis, we also used a maximum likelihood
method outlined by KF93
to measure the extragalactic ionizing
background in a manner that avoids binning of the data.  One constructs a
likelihood function of the form
\begin{equation}
L= \prod_{a}{\it f}(N_{a},z_{a}) \prod_{Q} exp [ - \int^{z^{Q}_{max}}
_{z^{Q}_{min}} dz \int^{\infty}_{N^{Q}_{min}} {\it f}(N,z) dN ],
\end{equation}
where  the subscripts $a$ and $Q$ refer to absorbers and QSOs and where
${\it f}(N,z)$ is the standard equation for the distribution of 
Lyman $\alpha$ absorbers in column density and redshift,
\begin{equation}
{\it f}(N,z) = AN^{-\beta}(1+z)^{\gamma}[1+\omega(z)]^{-(\beta-1)}.
\end{equation}

The parameter $\omega$ is defined as above, but here, the normalization
in terms of ${\cal A}_{0}$ in Equation~\ref{eq:dndz} is given by
${\cal A}_{0}(N_{lim}/N_{0})^{\beta-1}(\beta-1)^{-1}$.  With the exception
of the case in which a variable threshold is used,  $N_{min}$ for each
QSO is the column density which, according to the curve-of-growth
adopted (see KF93), corresponds to an equivalent width of 0.32 $\AA$,
2.62 $\times$ 10$^{14}$ cm$^{-2}$.  

Instead of using the method outlined by KF93 whereby the parameters
$A$,$\beta$,$\gamma$, and $J(\nu_{0})$ are all found by minimizing -ln(L)
where L is given by the likelihood function above, 
we chose to take the parameter $\gamma$ from
a separate maximum likelihood solution to Equation~\ref{eq:dndz} (see
Paper I.)  Since our spectra are more highly blended than the
low redshift data used by KF93, we choose not to determine $\beta$
directly from our data
using line equivalent widths and the curve-of-growth and 
instead adopt a value found from high resolution
spectra.
As described in the previous section,
we take $\beta$ to be 1.46 (Hu et al. (1995) 
and solve for $A$ by requiring
${\it f}$(N,z) to give the observed number of lines in the regions of
the QSO spectra unaffected by the proximity effect.

We ran two tests on this set of algorithms.  The first of these was to
attempt to reproduce the results of KF93 with the dataset they used from
Bahcall et al. (1993) .  Next, we used a high redshift subsample 
of our complete dataset, 
the DB96 sample, to compare the results of the maximum likelihood analysis
and the BDO analysis to each other and to independent checks on these
values (B94, Giallongo et al. 1996).

We were able to reproduce the results of KF93.  Using their Sample 2,
the Bahcall et al. (1993) sample 
minus one BAL QSO, PG 0043+039, we obtain ($\gamma$, $\beta$, log(A)) = 
(0.23, 1.47, 7.74) and log[$J(\nu_{0})$]=-23.0$^{+0.7}_{-0.6}$ 
for $b= 35$ km s$^{-1}$.  These agree with the values they find,
($\gamma$, $\beta$, log$(A)$)= (0.21, 1.48, 7.74), and the errors in these
values, $\sigma_{\gamma} \sim 0.06$ $\sigma_{\beta} \sim 0.05$, and
$\sigma_{log(A)} \sim 0.1$.  Their result for log[$J(\nu_{0})$] for this sample
is -23.3$^{+ 0.7}_{- 0.5}$.

The high redshift subsample we created consisted of 518 lines from the 15 
objects from DB96 that do not show associated absorption.  The QSOs have
redshifts between 2.52 and 3.38.
Using our maximum likelihood program to solve for the Ly$\alpha$ forest
statistics, we find $\gamma= 1.926 \pm 0.656$, and log$(A)=
7.03$ for $N_{min}$=2.6 $\times$ 10$^{14}$ cm$^{-2}$ and 
$\beta$= 1.46. 
This subsample does give similar results in the BDO and the 
maximum likelihood cases, log[$J(\nu_{0})$]=$-21.40^{+1.10}_{-0.69}$ and
log[$J(\nu_{0})$]=$-21.58^{+0.30}_{-0.23}$, respectively. (See rows 
1 and 2 of Table~\ref{table-jnu}.)  These values agree
well with the Giallongo et al. (1996) result of log[$J(\nu_{0})$]=$-21.30 \pm 0.7$
for z=1.7-4.1.

The software we used for the maximum likelihood analysis uses all regions
of the QSO spectra between $z_{min}$, specified by the spectral coverage or
by Ly$\beta$ emission, and $z_{max}$, specified by Ly$\alpha$ emission.
Though it does not count lines associated with identified metal line systems,
it does not exclude the regions of the spectrum where these lines lie.  To
ensure that this does not have a significant effect on our resultant solution
for the background, we tested a program that does exclude regions of the
spectra in the same way that our BDO-style software does.  The change in the
result was indeed insignificant; but taking these excluded spectral regions
into account and binning the data in the same way the BDO-style software
does brings the maximum likelihood and the BDO method results into excellent
agreement.

Figure~\ref{fig:like} plots the log of the ratio of the likelihood function to
the maximum value versus log[$J(\nu_{0})$]; and 
Figure~\ref{fig:dndxl} plots the coevolving number distribution of
Ly$\alpha$ lines
with respect to $\omega$ just as in Figure~\ref{fig:dndx}.
The results of this analysis are also summarized in Table~\ref{table-jnu}
and discussed further in Section~\ref{sec-disn}.

\subsection{Systemic QSO Redshifts}
\label{sec-zs}
One of the major uncertainties in the proximity effect analysis is
in the systemic redshifts of the QSOs.
If the true redshift of a QSO is higher than the value used in the
analysis, any given cloud is further away from the QSO than assumed.  Hence,
the influence of the QSO at this cloud is less than inferred and the value
of $J(\nu_{0})$ in reality is lower than the one derived. 

For the data presented in Figure~\ref{fig:nlz}, an average of several
cursor settings at the peak of the emission line was used
to determine the line centers.  More detailed fits were not done as our
purpose lies mainly in determining if any gross shifts between Ly$\alpha$
and the Balmer lines/[OIII]/Mg II exist for our data; but we found
no significant difference between this method and making Gaussian
fits to the upper 50\% of the emission line profiles.
 
Ly$\alpha$ redshifts were measured from the absorption line spectra 
when the entire Ly$\alpha$ profile was observed, in the same way as 
was done for the Balmer, [OIII], and Mg II lines.
Table~\ref{table-z} lists the adopted best redshift value for each
emission line for each object
supplementing our measurements with measurements from the literature.

Laor et al. (1994) and Laor et al. (1995) found, from a sample
of 13 QSO spectra from the Faint Object Spectrograph on Hubble Space Telescope
between redshifts of $z \sim 0.16$ and  $z \sim 2.0$,
average velocity
shifts between [OIII] $\lambda$5007 and Ly$\alpha$, Mg II, and H$\beta$
of $200 \pm 150$ km s$^{-1}$, $-85 \pm 130$ km s$^{-1}$, and $-75 \pm
110$ km s$^{-1}$,
respectively.  This agrees with the Corbin \& Boroson (1996) result
for 48 objects with $0.03 < z < 0.77$.
They found mean [OIII]-Ly$\alpha$ and [OIII]-H$\beta$ shifts of 
$191 \pm 101$ km s$^{-1}$ and $-75 \pm 57$ km s$^{-1}$. 
Thus, Ly$\alpha$ is blueshifted with respect to
[OIII] by $\sim$200 km s$^{-1}$, while Mg II and H$\beta$ are 
marginally redshifted with respect to [OIII].
Tytler \& Fan (1992) find a mean [OIII]-H$\beta$ shift of
$-15 \pm 37$ km s$^{-1}$ from 8 QSOs  with redshifts between $\sim$0.3 and
$\sim$0.6 and conclude that both Balmer lines and narrow forbidden lines
give redshifts within 100 km s$^{-1}$ or less of the QSO systemic
redshift.  They then find a {\it blueshift} of  Mg II with respect to
[OIII]/H$\beta$ for 100 QSOs
of $101 \pm 47$ km s$^{-1}$ which they use as a secondary systemic
redshift zero point in their analysis of a large QSO sample.  The
magnitude of the blueshift of Ly$\alpha$ with respect to [OIII]/H$\beta$
that they derive is $172 \pm 17$ km s$^{-1}$.
The data of Nishihara et al. (1997) for five QSOs at $z \sim 1.5$
show a negligible redshift of Mg II
with respect to [OIII], $31 \pm 411$ km s$^{-1}$.  However these five
objects show a somewhat larger redshift of 
H$\beta$ with respect to [OIII] $\lambda$5007,
equalling $260 \pm 522$ km s$^{-1}$,
consistent with the fact that these objects have high luminosities. 
M$^{\rm c}$Intosh et al. (1999b) use the near-infrared spectra of 
QSOs at 2.0 $\lesssim$ z $\lesssim$ 2.5 presented
in M$^{\rm c}$Intosh et al. (1999a) to examine the redshift differences 
between [OIII] and H$\beta$. They supplement their data with data from
the literature to measure the redshift differences between [OIII] and Mg II.
They find that on average, H$\beta$ is redshifted relative to
[OIII] by 520 $\pm$ 80 km s$^{-1}$ for 21 of their sample objects,
while Mg II lies within 50 km s$^{-1}$
of the redshift of [OIII] for 12 sample objects. 

For our sample, we find that Ly$\alpha$ 
is blueshifted with respect to [OIII] $\lambda$5007 by 
$382 \pm 1160$ km s$^{-1}$ for 19 QSOs.
Mg II emission is blueshifted 
by an average of $338 \pm 901$ km s$^{-1}$ with respect
to [OIII] on the basis of seven measurements.
We find that H$\beta$ is redshifted by  $642 \pm 740$ km s$^{-1}$
with respect to [OIII] on the basis of five measurements.
Including three H$\alpha$ redshifts listed in
Table~\ref{table-z} with these
H$\beta$ redshifts leads to a $507 \pm 615$ km s$^{-1}$
redshift of Balmer lines with respect to [OIII].  This
shift is larger than that discussed above for low reshift QSOs.  However,
it is consistent with the Nishihara et al. (1997) H$\beta$ shift for high
luminosity QSOs.  Combining our data with that of these authors, 
we find that Mg II is blueshifted with respect to
[OIII] by $184 \pm 735$ km s$^{-1}$.
Including the data of M$^{\rm c}$Intosh et al. (1999b) that is
not already in our sample
gives a blueshift of $95 \pm 603$ km s$^{-1}$. Similarly, combining our data
with that of Nishihara et al. (1997),
we find that H$\beta$ is redshifted with respect to [OIII] by $451 \pm 636$ 
km s$^{-1}$. After supplementing this combined data set with
the data of M$^{\rm c}$Intosh et al. (1999b), the redshift
becomes $379 \pm 516$ km s$^{-1}$.
Finally, combining the data of M$^{\rm c}$Intosh et al. (1999b) 
with ours gives a Ly$\alpha$ blueshift of $418 \pm 920$ km s$^{-1}$ with
respect to [OIII].

As has been noted in previous work, the standard error in the mean
velocity shifts is quite large, on the order of or exceeding the
value of the shift itself.  We estimate that the wavelength calibration
errors in our data contribute a $\sim$10-30 km s$^{-1}$ error in the
derived redshifts; and
the spread in different redshift measurements
of the same species (e.g. Balmer lines or [OIII] $\lambda$4959 and
$\lambda$5007) for the same object
is typically 100-200 km s$^{-1}$.  The observed spreads in the 
velocity differences of the Ly$\alpha$, Mg II, and Balmer emission lines
with respect to the QSO systemic redshifts
are much larger than this, indicating that it is intrinsic to the QSO 
population.
Figure~\ref{fig:zhist} shows histograms of the
emission line redshift differences between [OIII] and Ly$\alpha$, [OIII] and
Mg II, and [OIII] and Balmer lines.  Our results
are plotted with those of Laor et al. (1995) and of Nishihara et al.
(1997). Our
sample shows no well-defined mean [OIII]-Balmer shift, just
a large scatter in the measurements included.
Our sample also
shows a large range of [OIII]-Ly$\alpha$ and [OIII]-Mg II shifts with
no well-defined mean value.  Nonetheless, the mean trend is that
the [OIII]-Ly$\alpha$ shift is different from zero by 1.4$\sigma$ for our
data, 
less than the 3.5$\sigma$ significance found by Laor et al.
(1995).  The [OIII]-Balmer line shifts for both our data set and for our data
combined with that of Nishihara et al. (1997) are {\it more} significant,
2.7$\sigma$ and 2.8$\sigma$ respectively.
The [OIII]-Mg II shift is consistent with
zero in a mean sense, but with large scatter.
Thus, though better statistics are desirable, 
it seems that for these high redshift and relatively high luminosity
objects, Balmer lines are not good indicators of the QSO systemic redshift.
For the purposes of this study therefore,
we treat only the redshifts found from [OIII] $\lambda$5007 for 19 objects
in our sample and 
Mg II for 16 objects in our sample as 
systemic QSO redshifts.

\subsection{The HI Ionization Rate}
\label{sec-gamma}

The HI
ionization rate due to a source of UV flux is formally given by the equation: 
\begin{equation}
\Gamma=\int_{\nu_{0}}^{\infty} \frac{4 \pi J(\nu) \sigma_{HI}(\nu)}{h \nu} d\nu 
\; \; {\rm s^{-1}}.
\label{eq:gamint}
\end{equation}
The calculations of the mean intensity of the ionizing background to date
have made a critical assumption, namely that the spectrum of the background
and the spectra of the individual QSOs are identical. This allows the
expression $\omega= \Gamma^{Q}/\Gamma^{bg}$
to reduce 
to the ratio of the Lyman limit flux density of the QSO, $J^{Q}(\nu_{0})$, to that of 
the background, $J^{bg}(\nu_{0})$, for each line (BDO).
Since the IGM
reprocesses the radiation emitted from QSOs, this is not strictly
true (Miralda-Escud\'{e} \& Ostriker 1990, Madau 1991,1992, 
Meiksin \& Madau 1993, Haardt \& Madau 1996, Fardal et al. 1998).  
Furthermore, 
the value of $\Gamma^{bg}$ is of particular interest as it
can be used to infer the value of 
$\Omega_{b}$ by comparing the distribution of flux decrements in high resolution
QSO spectra to Lyman $\alpha$ forest simulations (Rauch et al. 1997). 
Therefore, we repeat the
standard BDO analysis without making this assumption, ie. using 
$\omega= \Gamma^{Q}/\Gamma^{bg}$
and solving for the HI
ionization rate from the metagalactic background radiation. The ionization 
rate for each QSO was calculated using Equation~\ref{eq:gamint}, where 
$\sigma_{HI}(\nu)=6.3 \times 10^{-18} (\frac{\nu_{0}}{\nu})^{3}$ cm$^{2}$ and where
$J^{Q}(\nu)=J^{Q}(\nu_{0})(\frac{\nu}{\nu_{0}})^{-\alpha}$.  For each QSO, $J^{Q}(\nu_{0})$ 
is the same value
used in the standard analysis used to solve for $J^{bg}(\nu_{0})$, 
and $\alpha$ is given in Table~\ref{table-specphotfinal}.
For some objects, no $\alpha$ listed in this table and a value of 0.46 was used, as
described in Section~\ref{sec-pespecphot}.
As before,
the best value will be the one that gives the lowest $\chi^{2}$ between the
model with $\beta$=1.46 and the binned data.
We use the
narrow line redshifts for each QSO discussed above and add 400 km s$^{-1}$
to each QSO redshift measured from the Lyman $\alpha$ emission line.  

Haardt \& Madau (1996) present a Gaussian fit to their model for the
evolution of $\Gamma$ with redshift,
\begin{equation}
\Gamma= A(1+z)^{B}exp[-(z-z_{c})^{2}/S]
\end{equation}
that agrees with their detailed model for the background to within 10\% over
the range $0 < z < 5$.  
The best fit parameters they derive for $q_{0}$=0.5 are $A$=6.7 $\times$ 10$^{-13}$
s$^{-1}$, $B$=0.43, $z_{c}$=2.30, and $S$=1.95. Fardal et al. (1998) fit their
model for the background with the parameter sets $A$=5.6 $\times$ 10$^{-13}$ s$^{-1}$,
$B$=0.60, $z_{c}$=2.22 , and $S$=1.90  and
$A$=1.26 $\times$ 10$^{-12}$ s$^{-1}$,
$B$=0.58, $z_{c}$=2.77 , and $S$=2.38 for the Q1 and Q2 luminosity functions,
of Pei (1995) respectively. Incorporating this expression for $\Gamma(z)$
with these three different sets of parameters into the BDO style analysis
allows us to determine which of these models fits our data best.  The results
are listed in Table~\ref{table-hm} and are discussed in greater depth 
below in Section~\ref{sec-disn}.

\section{Simulations and the Curve of Growth}
\label{sec-cog}

Simulated Lyman $\alpha$ forest spectra for the DB96 sample only were
produced using the software described in that paper.  The simulation input
$\gamma$ was changed slightly to reflect the maximum likelihood 
value found by the software used in the analysis described in Paper I.
The normalization was chosen to give matching amounts of total absorption
in the real and simulated spectra. 
The parameters used were $\gamma= 2.069$, ${\cal A}_{0}= 4.835$, $\beta= 1.46$,
log(N$_{\rm HI_{min}}$) = 13.0, log(N$_{\rm HI_{max}}$) = 16.0, $<b>$= 28.0 km s$^{-1}$,
$\sigma_{b}$= 10.0 km s$^{-1}$, and $b_{\rm cut}$= 20.0 km s$^{-1}$. 

The proximity effect was included in these simulations by simply 
modifying each cloud's column density according to equations~\ref{eq:column}
and ~\ref{eq:omega}.  The value of log[$J(\nu_{0})$] from the BDO type analysis
on the DB96 sample is -21.40$^{+1.1}_{-0.69}$.
Values of  -19.0, -20.0, -21.3, -22.0, and -23.0 for log[$J(\nu_{0})$]
were input and the
analyses described above were used to recover that $J(\nu_{0})$.  Two examples
of the simulated spectra are shown in Figure~\ref{fig:simspec}.

The analysis considers all lines above a fixed equivalent width threshold
of 0.32 $\AA$.  Thus, as the column densities of lines are modified by
the QSO flux from their expected values in the absence of the proximity
effect, the equivalent widths of the lines will change according to the
curve-of-growth.  If a line is saturated, changing its column density
will have little effect on its equivalent width, since it lies on the flat
part of the curve-of-growth where $W \propto \sqrt{log(N)}$.  This will
mean that for a given equivalent width cutoff in the data, this line will 
not drop out of the sample as the proximity effect is turned on in the
simulations.  Since the line deficit will be less than expected for a given
input value of $J(\nu_{0})$, the proximity effect will appear less pronounced and
the true $J(\nu_{0})$ will be overestimated.  We found this to be the case from
our simulations. As Figure~\ref{fig:simdndx} illustrates and 
Table~\ref{table-sim} summarizes, 
though the values of $J(\nu_{0})$ recovered
from the simulated data were usually consistent with the input values within
the 1$\sigma$ confidence limits, 
they were systematically larger than the input values by up to
a factor of 3.  
The largest input values of log[$J(\nu_{0})$], -19.0 and -20.0, 
give the largest discrepancy between this input value and the log[$J(\nu_{0})$]
recovered from the BDO analysis performed on the simulated spectra.
The smallest input value of log[$J(\nu_{0})$], -23.0, gives 
the smallest discrepancy between the input and recovered values.
However, the 1$\sigma$ confidence limits on this fit are also relatively small,
making it the only trial which does not recover the input log[$J(\nu_{0})$] to within
those limits. 

To demonstrate the effect, Figure~\ref{fig:cog}  compares the simulated line 
equivalent widths
with and without the proximity effect included. 
The column density of each line from 
the simulated spectra line lists with no proximity effect were modified 
according to  equations~\ref{eq:column} and ~\ref{eq:omega}.  Figures 12(a-e) 
plot the nonproximity effect rest equivalent width $W_{no-PE}$
versus the ratio
of the proximity effect and nonproximity effect equivalent widths,
$W_{PE}/W_{no-PE}$.  
The  solid line delineates the detection  threshold for the lines in the
list for which the proximity effect is included, 
$W_{PE}$= 0.32 $\AA$. Absorption lines that fall
above this line were not removed from the sample when the proximity effect
was turned on, while those below it disappeared.  
For a given set of QSOs with fixed Lyman limit lumosities, such
as this one,
the proximity effect signature in their spectra will
become less pronounced as the ambient UV background increases.
Therefore, as log[$J(\nu_{0})$]
increases from -23.0 to -19.0, the
magnitude of the proximity effect decreases, and 
the proximity effect line list differs less and less from the nonproximity effect line list.

\section{Results and Discussion}
\label{sec-disn}

Table~\ref{table-jnu} lists the best fit values of $J(\nu_{0})$ found for
various subsamples of this dataset using both the canonical BDO 
and the maximum likelihood methods. For the BDO method, 
the 1$\sigma$ confidence
limits are found from a $\Delta \chi^{2}$ of 8.18 for 7 degrees of freedom.  
The maximum likelihood method 1$\sigma$ confidence limits derive from
the fact that ln$(L/L_{max})$ is distributed as $\chi^{2}/2$.   
The total sample
consisting of 74 QSOs with all QSO redshifts based on the Ly$\alpha$ emission
line gives a best fit value of log[$J(\nu_{0})$] of -20.90$^{+0.61}_{-0.48}$ 
for the BDO analysis and -20.83$^{+0.23}_{-0.20}$ for the maximum likelihood
analysis.

As the results in  Table~\ref{table-jnu} demonstrate,  
using narrow line redshifts for 35 of the 74
QSOs for which they have been directly measured and Ly$\alpha$ redshifts 
for the rest does not change the 
result. However, when 400 km s$^{-1}$ is added to
the Ly$\alpha$ redshifts of the objects with no measured narrow line redshift,
a value for log[$J(\nu_{0})$] of -21.15$^{+0.17}_{-0.43}$ is derived using
the BDO method and log[$J(\nu_{0})$]=-21.17$^{+0.19}_{-0.15}$ is found using
the maximum likelihood method. Recall that the mean blueshift of
Ly$\alpha$ with respect to [OIII] for the 19 objects in this paper
with [OIII] $\lambda$5007 measurements was found to
be $\sim$400 km s$^{-1}$.  This decrease in the mean intensity of the
background derived when larger QSO redshifts are used is to be expected.
(cf. Section~\ref{sec-zs})  
Because this measurement of the background accounts for the systematic
blueshift of the Ly$\alpha$ emission line with respect to the systemic
redshift of each QSO, we consider it to be our best estimate for the mean
intensity of the background at the Lyman limit.

These measurements have been made, however, using a photoionization model
with somewhat unrealistic assumptions, particularly that Ly$\alpha$ 
absorbers are isothermal and are composed of pure hydrogen.
For clouds  with a primordial
He abundance and which are in thermal and ionization equilibrium, 
Using CLOUDY to model the ionization state of absorbers with a metal
adundance of 10$^{-2}$ solar (Cowie et al. 1995, Tytler \& Fan 1994)
as a function of $\omega$, we find 
that the neutral fraction, $\chi$, is proportional
to $(1+\omega)^{-1.21}$.  
This implies that 
\begin{equation}
\frac{d{\cal N}}{dz}= {\cal A}_{0}(1+z)^{\gamma}[1+\omega(z)]^{-1.21(\beta-1)}
\end{equation}
In this scenario, the optimal value found for 
log[$J(\nu_{0})$] is -21.10$^{+0.53}_{-0.28}$.  This value is 
marginally larger than the value discussed above, found under the assumption
of absorbers composed of pure hydrogen; but it is not 
significantly different, so 
we conclude that the absence of metals in the BDO model has not drastically
affected our measurement of the background.

It is worth noting that 16 objects in our sample of objects with
no associated absorption show
evidence for damped Ly$\alpha$ absorption: 0058+019, 0100+130, 0334-204,
0913+072, 0938+119, 0952+338, 0955+472, 1009+299, 1017+280, 1215+333,
1247+267, 1548+092, 1946+770, 2126-158, 2233+131, and 2320+079.
The dust in these systems could
cause the intrinsic QSO fluxes to be underestimated. This in turn can cause
log[$J(\nu_{0})$] to be underestimated by up to a factor of 3,
in addition to the sources of error discussed above (Srianand \& Khare 1996).
Only six of these objects, 0334-204, 0938+119, 0955+472, 1215+333, 2126-158, 
and 2233+131, appear in our low
luminosity subsample, suggesting that this subsample is not preferentially
heavily dust-obscured.  Nevertheless,  the BDO analysis was performed on
all 16 objects exhibiting damped  Ly$\alpha$ systems; and  found the best fit
value for log[$J(\nu_{0})$] to be -21.45$^{+0.40}_{-0.53}$, a factor of
1.9$^{+8.1}_{-1.6}$
lower than the value obtained for the sample as a whole. 
This does not allow us to say anything
significant about the presence or absence of dust, so we will neglect its
influence.

Dividing our line sample into subsamples of high ($z > 2.5$) and low ($z < 2.5$)
redshift lines, we find marginal evidence for evolution in the intensity of the
background, namely that the maximum likelihood background intensity 
is lower by a factor of
about 1.9$^{+3.9}_{-1.4}$ at lower redshift. 
The BDO results corroborate this, but with larger uncertainties.
The factor by which $J(\nu_{0})$ is found to be lower at lower redshifts is
2.5$^{+27.7}_{-2.2}$.
Gravitational lensing  could
mimic a trend with redshift with about the same order of magnitude, if 
the high redshift subsample contains a significant number of unknown lenses.
However, Figure~\ref{fig:zlum} suggests little if any trend for high luminosity
objects to exist at high redshifts in our sample; and the 
results of Section~\ref{sec-npr} indicate that the high
luminosity objects do show a somewhat stronger proximity effect despite the
fact that the measured background at high redshift appears to be
{\it higher}.
No other studies have found this evidence of redshift evolution in the
background, so we regard it as tentative; and note that
it will be interesting to  see in future work if this trend can be shown to be
real and if it extends smoothly
to the low values of $J(\nu_{0})$ found at redshifts less than 1.5.

Since we find high luminosity objects do not exist preferentially at high
redshift in our sample,
a simple test can be done to determine whether or not 
there is a significant number of
lensed objects in our sample.
If the high luminosity QSOs are indeed intrinsically more luminous,
and the proximity effect is a purely photoionization-driven phenomenon,
these objects should show a more prominent proximity effect.
The results of Section~\ref{sec-npr} suggest this is the case.  
However, in the analysis, this  larger line deficit is normalized  to
the higher Lyman limit luminosities of this subsample.
Therefore,
one expects these objects, when analyzed as a separate subsample, to yield a 
value of $J(\nu_{0})$ that is consistent with that found for low luminosity objects
if the values of the QSO fluxes are not in error due to lensing.
If the high luminosity QSOs, or a subset of them,
are lensed objects, then they are not necessarily intrinsically more luminous
than the low luminosity QSOs.  In this case, the influence of the lensed
objects on the surrounding IGM will be overestimated
and given the observed line deficit, the background will also be
overestimated. 
Table~\ref{table-jnu} lists the results obtained for the
high and low luminosity subsamples of our data set.  The values obtained
for these subsamples are equal within the uncertainties. This is consistent with
there being no significant effects from gravitational lensing in our
sample.  

\subsection{HI Ionization Rate}
We tested a range of values for $\Gamma$, the HI ionization rate,
using our data. The constant value
found to fit the data the best is  1.9$^{+1.2}_{-1.0}$ $\times$ 10$^{-12}$ s$^{-1}$.
This value is in reasonable agreement 
with that predicted by the QSO-dominated model of 
Haardt \& Madau (1996) at this
redshift, 1.0 $\times$ 10$^{-12}$ s$^{-1}$ ($q_{0}$=0.5).  
Using Equation~\ref{eq:gamint} and
$J^{Q}(\nu)=J^{Q}(\nu_{0})(\frac{\nu}{\nu_{0}})^{-\alpha}$, and assuming
global QSO spectral indicies of
0, 1.5, and 2, the ionization rate found from our data corresponds to
log[$J(\nu_{0})$]= -21.34, -21.17, and -21.12, respectively.

The parameter set ($A, B, z_{c}, S$) found to give the best fit to the data is
that of Fardal et al. (1998) for the Q2 luminosity function 
(1.2 $\times$ 10$^{-12}$ s$^{-1}$, 0.58, 2.38, 2.77) which, for a 
redshift of 2.9 yields an
ionization rate of 2.7 $\times$ 10$^{-12}$ s$^{-1}$, in good agreement with
our solution,
and within a factor of $\sim$3 of the Haardt \& Madau result.
Thus, we conclude
that a significant contribution to the ionizing background from stellar
UV emission is not required at this redshift.

\subsection{Curve-of-Growth and Other Systematics}
On the basis of a curve-of-growth argument,
one might expect that weak lines would
show a more prominent proximity effect than strong lines.  We have compared
the results obtained for a constant equivalent width threshold of 0.32 $\AA$
with that obtained for lines with 0.16 $\AA < W <$ 0.32 $\AA$. 
Instead of finding
a more pronounced proximity effect for the weak lines, we find a less 
significant deficit of lines within 1.5 $h^{-1}$ Mpc of the QSOs.
This deficit is 4.0$\sigma$, versus  
5.5$\sigma$ for lines with $W >$ 0.32 $\AA$.  As Table~\ref{table-jnu} lists,
the value of log[$J(\nu_{0})$]
recovered from these weak lines is correspondingly higher than that found
using strong lines,  -20.45$^{+0.37}_{-0.90}$ versus -21.15$^{+0.17}_{-0.43}$.
Cooke et al. (1997) point out that 
this could be the result of a higher degree
of blending of weaker lines compared to strong ones in crowded spectral regions.
The background flux measurement will be an overestimate
because blending will cause fewer
individual lines to be resolved further from the QSO.  Because the reduction
in line density near the QSO will  work to reduce line blending, 
the overall effect
of line blending will be to suppress the true magnitude of the proximity effect 
causing $J(\nu_{0})$ to be overestimated, by a factor of 4.5 in this case.  It
is difficult to ascertain whether this effect is as strong for 
lines with $W >$ 0.32 $\AA$ or whether the curve-of-growth effect
discussed in Section~\ref{sec-cog} 
which also causes $J(\nu_{0})$ to be overestimated, 
is more important.  
We expect that for lines 
with $W >$ 0.32 $\AA$, the effects of blending are reduced somewhat, while
the curve-of-growth effects will remain a factor.

We have addressed many of the systematics which could
possibly have affected our analysis.   A treatment of the
QSO systemic redshifts was integrated directly into our analysis and
was found to influence the $J(\nu_{0})$ found by up to a factor of 
$\sim$2.
Other effects, such as the influences of metals and dust, which can
cause $J(\nu_{0})$ to be underestimated, and the influences of lensing,
line blending, and curve-of-growth effects, which can cause $J(\nu_{0})$
to be overestimated, were treated after the fact in an attempt to understand
the magnitude of their effects on the value of $J(\nu_{0})$ derived.
The CLOUDY simulations discussed above indicate that allowing for 
an absorber metal abundance of 10$^{-2}$ solar has little effect
on the value of $J(\nu_{0})$ found from the data. Dust in intervening
absorption systems may have affected our result. Though we were unable to
quantify this effect with high confidence, it could be on the order of
a factor of 2.
We assert that QSO flux amplification due to lensing has not
significantly biassed our result; and we attempt to
minimize the effect of blending discussed above by using only lines with
$W >$ 0.32 $\AA$.
Our result may be susceptible to the
curve-of-growth effect we addressed through the simulations in 
Section~\ref{sec-cog}.  In those simulations, we found that the
discrepancy between in the input and recovered values of $J(\nu_{0})$
depended upong the input value of $J(\nu_{0})$ itself.  The
magnitude of the discrepancy corresponding to the $J(\nu_{0})$
we found from the data was a factor of $\sim$2.  We therefore suspect
that if our result, log[$J(\nu_{0})$]=-21.15$^{+0.17}_{-0.43}$, 
is systematically biased in any
way, it is an overestimate of the true background and could be
in error by up to a factor of 2; though this could be balanced
somewhat by systematic error due to dust, which works in the 
opposite direction.

\subsection{Comparison with Previous Measurements}

Our value for $J(\nu_{0})$ agrees well with other measurements
at similar redshift, with the exception of those of B94 and
Fern\'{a}ndez-Soto et al. (1995) who both derive values four times 
larger than our best value for $J(\nu_{0})$, $\sim$3 $\times$ 10$^{-21}$
ergs s$^{-1}$ cm$^{-2}$ Hz$^{-1}$ sr$^{-1}$.  The measurement of B94 does not
take into account QSO systemic redshifts, but she notes that if they are
blueshifted with respect to Ly$\alpha$ by 1000 km s$^{-1}$, this would
lower the derived value of $J(\nu_{0})$ by a factor of 3, bringing it into
reasonable
agreement with our result.  The Fern\'{a}ndez-Soto et al. (1995)
value is derived from 3 QSO spectra showing a proximity effect due to
foreground QSOs.  
These authors are not able to place an upper limit on their 
measurement, but our value of 7.0 $\times$ 10$^{-22}$ ergs s$^{-1}$ cm$^{-2}$ Hz$^{-1}$
sr$^{-1}$  for $J(\nu_{0})$ is consistent with their lower limit of
1.6 $\times$ 10$^{-22}$ ergs s$^{-1}$ cm$^{-2}$ Hz$^{-1}$ sr$^{-1}$.  
In fact, when these authors examine the proximity effect in a single
QSO spectrum due to the background z$\sim$2 QSO itself, they derive a value for 
$J(\nu_{0})$ of  7.9$^{+23.}_{-6.0}$ $\times$ 10$^{-22}$ ergs s$^{-1}$ cm$^{-2}$ Hz$^{-1}$
sr$^{-1}$,
which brings their estimate into better agreement
with our values for
our total sample and for our low redshift subsample within their large errors.
Direct measurements of the background at redshifts $\sim$3-3.5 have been
made using long-slit spectroscopy of fields containing optically thick
Ly$\alpha$ absorbers in efforts to detect fluorescent emission the absorbers
produce from the ionizing radiation field incident upon them (Lowenthal
et al. 1990, Mart\'{i}nez-Gonz\'{a}lez et al. 1995). 
Recent Keck telescope observations by Bunker et al. (1998) at 2.5 $<$ z $<$ 4.1 
have achieved a factor of 2-10 higher sensitivity and place a firmer direct
limit on the background than previous work.
Their null signal
in a 90-minute integration with a 3$\arcmin$ slit sets an upper limit
on $J(\nu_{0})$ of 2 $\times$ 10$^{-21}$ ergs s$^{-1}$ cm$^{-2}$ Hz$^{-1}$ sr$^{-1}$.

Cooke et al. (1997) claim that the value for the background at z$\sim$4  
is between their value of 8.0$^{+8.0}_{-4.0}$ $\times$ 10$^{-22}$
ergs s$^{-1}$ cm$^{-2}$ Hz$^{-1}$ sr$^{-1}$
and that of Williger et al. (1994), 1.0-3.0 $\times$ 10$^{-22}$
ergs s$^{-1}$ cm$^{-2}$ Hz$^{-1}$ sr$^{-1}$.  Our best value of $J(\nu_{0})$
at z$\sim$3,
7.0 $\times$ 10$^{-22}$ ergs s$^{-1}$ cm$^{-2}$ Hz$^{-1}$ sr$^{-1}$, is in agreement
with this, although within the uncertainty there is 
an allowance for the background to
decrease as z approaches 4. 

Table~\ref{table-lit} lists these various measurements of $J(\nu_{0})$ in the
literature as well as the Kulkarni \& Fall (1993) measurement at
z$\sim$0.5.  Figure~\ref{fig:sum} also summarizes the 
literature measurements of
$J(\nu_{0})$ from z$\sim$0.5 to z=4.5.

The solid curves in Figure~\ref{fig:sum} 
delineate the evolution of the mean background
intensity as a function of redshift for
global background source spectral indicies between 0 and 2, derived from the 
Haardt \& Madau (1996) model for the HI photoionization rate as a
function of redshift discussed in Section~\ref{sec-gamma}.  
Over 90\% of our sample QSO redshifts lie within the FWHM of
the Gaussian in the  Haardt \& Madau (1996) expression using
their best fit parameters.
At these redshifts, the Haardt \& Madau (1996) curves in Figure~\ref{fig:sum}
are turning over.  Nonetheless, for comparison with previous work (B94 and references
therein),
we investigate a power law
redshift dependence of the background intensity:
\begin{equation}
J(\nu_{0},z)=J(\nu_{0},0)(1+z)^{j}.
\end{equation}
Using the BDO method,
we executed a crude grid search in an attempt to constrain
the power law index and normalization of this power law.
The lowest $\chi^{2}$ (3.86) between the binned data and the BDO photoionization
model for a power law background was achieved by ($j$, log[$J(\nu_{0},0)$])=
(5.12, -23.97),
shown by a dashed line in Figure~\ref{fig:pl}.   
Extending this solution to low redshift gives log[$J(\nu_{0}),0.5)$]=-23.0,
in good agreement with the measurement of Kulkarni \& Fall (1993). 
The solution ($j$, log[$J(\nu_{0},0)$])= (-4.16, -18.76)  
gives the next lowest $\chi^{2}$ (4.91); and though
it also implies mean background intensities over four orders of magnitude
too high at low redshift, it traces the Haardt \& Madau model at
high redshift, giving log[$J(\nu_{0},4.5)$]=-21.8, in agreement with the
Willigher et al. (1994) measurement.  It is also shown by a dashed line
in Figure~\ref{fig:pl}.
Fitting parabolas to the regions near the  $\chi^{2}$ minima 
in both $j$ and log[$J(\nu_{0},0)$] gives the
error in each parameter for both of these solutions,
(5.12 $\pm$ 1.96, -23.97 $\pm$ 1.07) and (-4.16 $\pm$ 2.36, -18.76 $\pm$ 1.31).
B94 found a similarly large range of acceptable solutions: $-7 < j < 4$
and $-16.5 < $ log[$J(\nu_{0},0)$] $ < -23.0$. 
The large error bars on these fits indicate that the power 
law fit to the data is not well-constrained, due possibly to the fact
that the mean intensity of the background is turning over at the 
redshifts of our sample objects, as the Haardt \& Madau (1996) model
predicts.

\subsection{Comparison with Models for the Background}
Recent models of the ionizing background include not only the integrated
emission from QSOs but also 
a variety of other physical processes such as star formation in young, 
high redshift  galaxies and
attenuation  of UV photons by  Ly$\alpha$ absorbers 
and Lyman  limit systems (Miralda-Escud\'{e} \& Ostriker 1990, 
Madau 1991, 1992, Meiksin \& Madau 1993, Haardt \& Madau 1996,Fardal et al. 
1998). 
Madau \& Shull (1996) find that the production of 
metals in Ly$\alpha$ absorbers may also be a significant contributor
to the UV background at z $\approx$ 3.  Their contribution may be
up to 5 $\times$ 10$^{-22}$ ergs s$^{-1}$ cm$^{-2}$ Hz$^{-1}$ sr$^{-1}$,
assuming that the bulk of the metals in the Lyman $\alpha$ forest did not
form at z $>>$ 3, and assuming a Lyman continuum escape fraction, $f_{esc}$,
from a galaxy
of $\gtrsim$0.25.  They note, however, that $f_{esc}$ is essentially
unconstrained.

Past debate about how the space 
density of QSOs evolves at high redshift  (Koo \& Kron 1988;
Boyle et al. 1991; Irwin et al. 1991; Schmidt et al. 1991;
Warren et al. 1994; Kennefick et al. 1995) has
been clarified by recent radio surveys (Hook et al. 1995, 1998; Shaver
et al. 1996).  This work has demonstrated that the space density of
radio-loud QSOs decreases rapidly with redshift beyond z$\sim$3.
Since these surveys are unaffected by any presence of dust in the 
intervening IGM; and since they confirm
the behavior seen in optically selected surveys, they indicate that the
QSO population is truly  declining at high redshift.
Nevertheless, 
the discovery of QSOs with redshifts greater than 4 has brought better
agreement between the values of $J(\nu_{0})$ found via the proximity effect
and the values predicted by the models with QSOs primarily contributing
to the background (Madau 1992, Meiksin \& Madau 1993,
Haardt \& Madau 1996).  

Madau (1992) and Meiksin \& Madau (1993) estimate the QSO UV background
by integrating the QSO luminosity function (Boyle 1991) and including
the effects of attenuation by hydrogen in the IGM.   Their
estimates however, 1-3 $\times$ 10$^{-22}$ ergs s$^{-1}$ cm$^{-2}$ Hz$^{-1}$ sr$^{-1}$,
are still somewhat lower than the values derived in this paper.
The analysis of Haardt \& Madau (1996) takes into account the effects 
of various atomic processes
leading to the production of hydrogen-ionizing photons within Ly$\alpha$
absorbers and Lyman limits systems themselves.  They conclude that 
observed QSOs can account for number of ionizing photons required by the
proximity effect at z$\lesssim$4.  These authors find a 
value of log[$J(\nu_{0})$] equal to $\sim$-21.4 at
z=3, in good agreement with the  
value found in this paper at similar redshifts.  The solid lines in 
Figure~\ref{fig:sum}
show the results from the Haardt \& Madau (1996) 
model for two different values
of the global background source spectral index.  
The lower and upper curves show
the evolution of the background for indicies of 0 and 2 respectively.  
The literature measurements at redshifts between 1.7 and 3.6 agree well
with the model predictions.  The  z$\sim$0.5 measurement of
Kulkarni \& Fall (1993) falls below both model curves and the z=4.5  
measurement of Williger et al. (1994) falls above them. 

Madau, Haardt, \& Rees (1999) revisit the issue of the contribution of
high redshift, star-forming galaxies to the ionizing background in
light of recent work identifying such objects at 2 $<$ z $<$ 4.
(Steidel et al. 1996a,b; Madau et al. 1996; Lowenthal et al. 1997)
They calculate the critical photoionization rate necessary to reionize
a non-uniform intergalactic medium as a function of redshift.
This is compared to the expected contributions from QSOs and young,
star-forming galaxies.  There are uncertainties in estimating both of these.
The QSO luminosity function at z $>$ 4 must be extrapolated from that
at lower redshifts. There is also still some debate between theory and 
observations, eg. of the {\it Hubble Deep Field},
on the subject of a population of
low-luminosity QSOs (see Madau et al. 1999 and references therein) which
could cause the QSO luminosity function to steepen with lookback time,
making up for the dearth of observed objects at z $>$ 4.  The estimation of the 
galaxy contribution of ionizing photons is limited
by poor knowledge of luminosity function
of Lyman-break galaxies at z $>$ 4 as well as by the lack of constraints
upon $f_{esc}$. Nevertheless, the results are intriguing.  Assuming that
$f_{esc}$=0.5, Madau et al. (1999) find that the contribution of 
hydrogen-ionizing photons from star-forming galaxies  z $\sim$ 3
could exceed that from QSOs by a factor of more than 3.  
However, the QSO contribution at this redshift is sufficient, according
to these estimates, to ionize the IGM at this redshift. 
Deharveng et al. (1997) estimate a much lower $f_{esc}$ at z=0, 
less than 1\%,
based on  the local galaxy H$\alpha$ luminosity density. 
Furthermore,
Devriendt et al. (1998) make an independent estimation of the galaxy 
contribution to $J(\nu_{0})$ assuming damped Ly$\alpha$ systems to be the
progenitors of present day galaxies.  
Their semi-analytic models include a treatment of not only HI absorption
of Lyman limit photons 
in the intervening IGM, but also
of HI and dust absorption  in the interstellar medium of the 
photon-producing galaxies.  Their results show that constraining 
$f_{esc}$ in this way yields a much lower contribution to the UV background
from galaxies at z $>$ 2.  At z $\sim$ 2.5, their estimated QSO contribution
to $J(\nu_{0})$ is 3 orders of magnitude greater than that expected from galaxies.
Our measurement of $J(\nu_{0})$ is consistent with the UV background
being QSO dominated
in the models of both these authors and Haardt \& Madau (1996). 

In the models of Madau et al. (1999),
the scenario changes at z $\gtrsim$ 3.5.  At this redshift, the 
QSO contribution of ionizing photons falls below the critical limit needed
to photoionize the IGM; and
by z=5, it will fall short of the critical value by a factor of $\sim$4.
This implies that at high redshift, the contribution from young stars 
may become the dominant contributor to the background, with the caveat
that the space density of star-forming galaxies would have to be maintained
at the level observed at z $\approx$ 3, and that most of their UV
photons would have to be free to escape into the IGM.  The
Devriendt et al. (1998) models lead to the conclusion, however, that 
the galaxy contribution to the UV background is negligible at high
redshifts.

In conclusion,
the proximity effect data at present reflect that the UV background
at 2 $<$ z $<$ 4 is QSO dominated.  The discrepancies between this
model at low and high redshifts (Kulkarni \& Fall 1993, Williger et al. 1994)
indicate that the contribution to the background from galaxies may be of
larger relative importance.  We plan to undertake an analysis of the proximity
effect at low redshifts from a large sample of QSO spectra taken with the
Faint Object Spectrograph on the Hubble Space Telescope to place better
constraints on the background at 0.5 $<$ z $<$ 2.  Further observations of
objects at z $>$ 4 are also of particular interest to this subject.

\acknowledgements

We extend thanks to the staff of the
Steward Observatory Bok Telescope for
their assistance with the observations, to T. Aldcroft and J. Shields 
for providing data, to C. Foltz and D. M$^{\rm c}$Intosh for helpful discussions,
to J. M$^{\rm c}$Dowell for use of his program COLDEN, 
and to G. Ferland and associates for making the program CLOUDY available
for general use.  We also thank S. Morris for a helpful referee report.
J. S. acknowledges the support of the National Science Foundation
Graduate Research Fellowship and the Zonta Foundation Amelia Earhart
Fellowship. 
J. B. acknowledges support from AST-9058510 and AST-9617060 of the 
National Science Foundation.
A. D.  acknowledges support from NASA Contract No. NAS8-39073 (ASC).
V. P. K. acknowledges partial support from an award from the William F. Lucas 
Foundation and the San Diego Astronomers' Association.
This research has made use of the NASA/IPAC Extragalactic Database (NED) 
which is operated by the Jet Propulsion Laboratory, California Institute
of Technology, under contract with the National Aeronautics and Space
Administration. 

\pagebreak

\pagebreak

\clearpage
\clearpage

\clearpage
\begin{figure*}
\epsscale{1.00}
\plotone{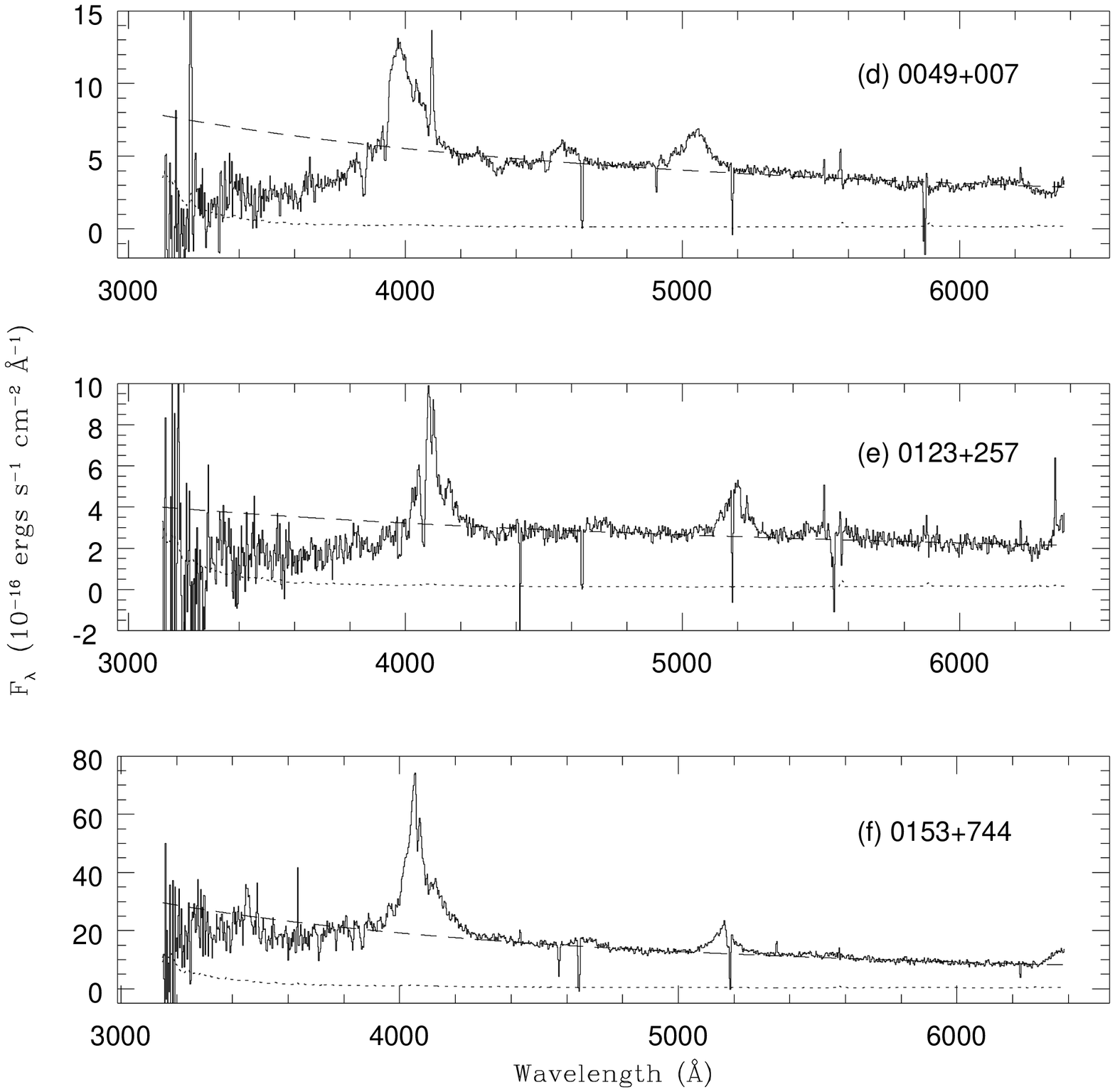}
\end{figure*}
\clearpage

\clearpage
\begin{figure*}
\epsscale{1.00}
\plotone{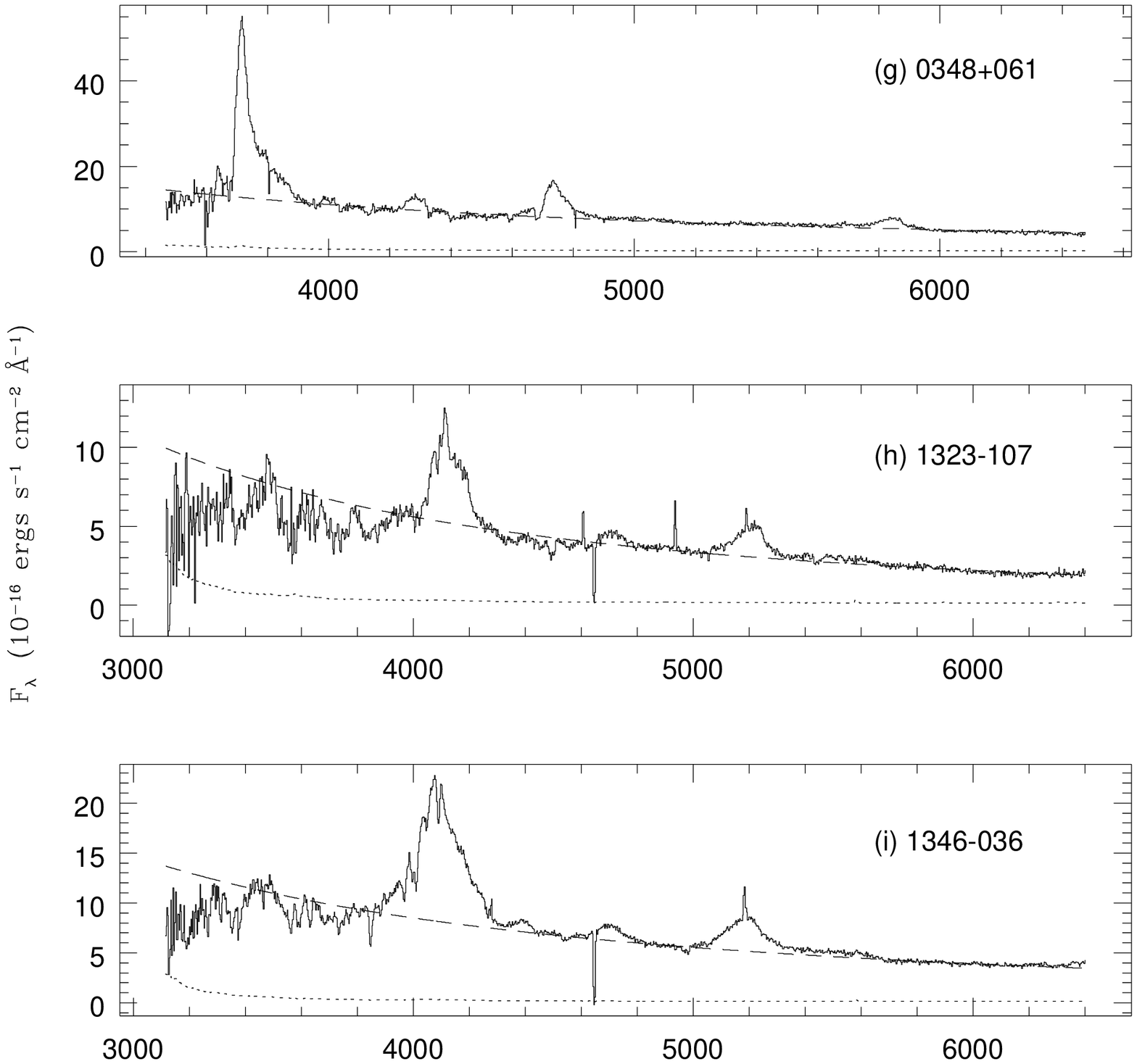}
\end{figure*}
\clearpage

\clearpage
\begin{figure*}
\epsscale{1.00}
\plotone{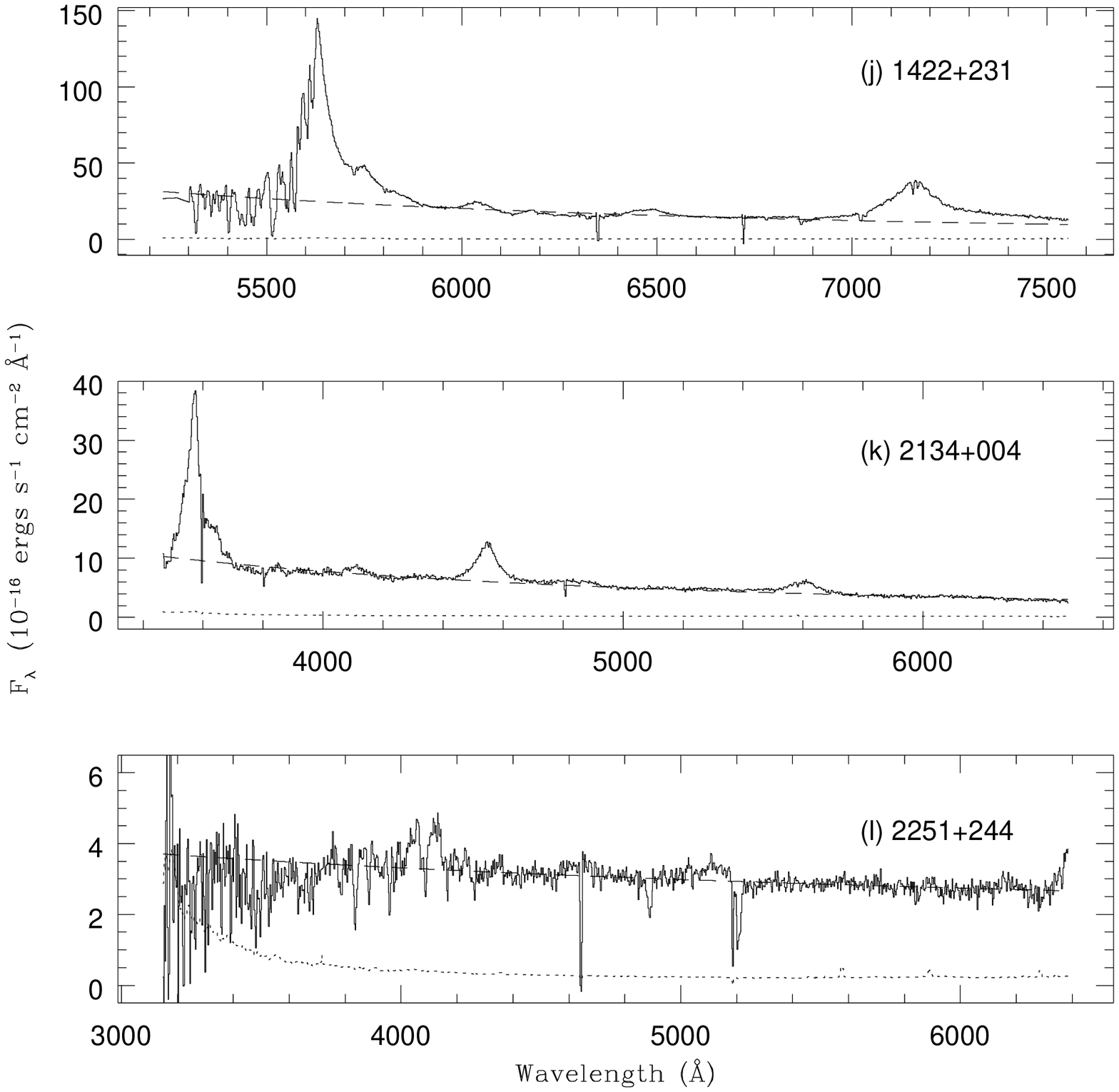}
\end{figure*}
\clearpage

\clearpage
\begin{figure*}
\epsscale{1.00}
\plotone{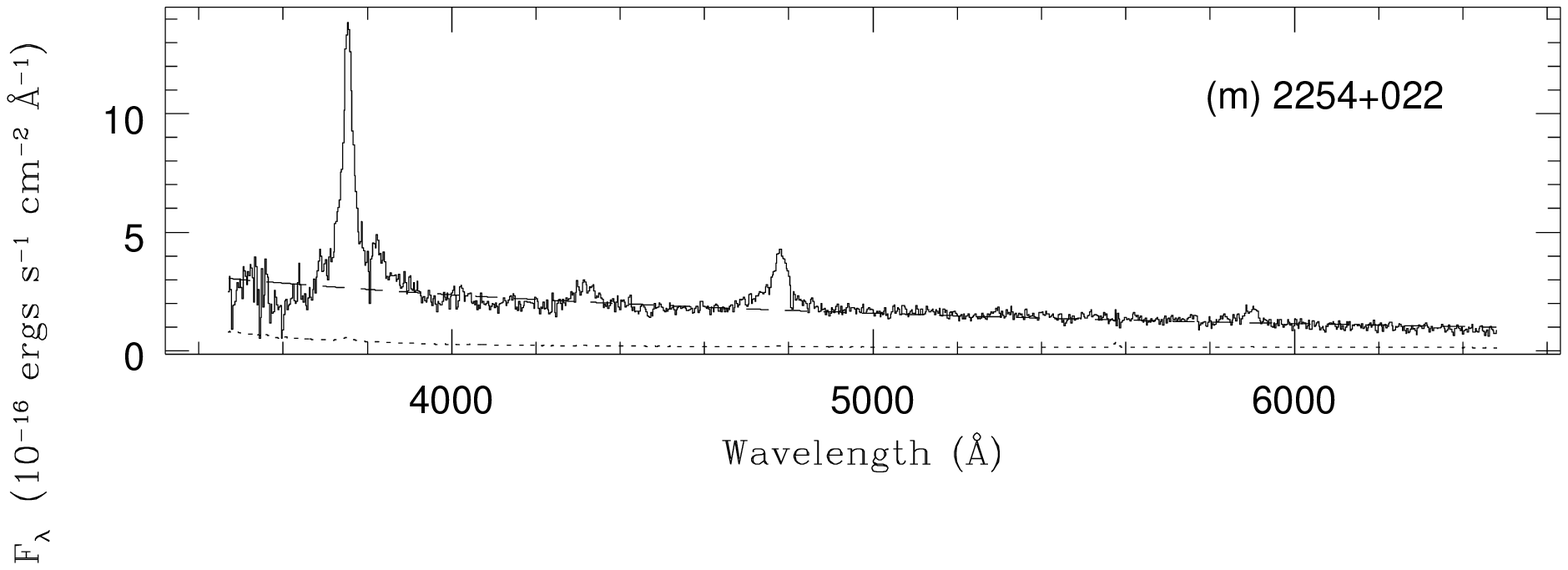}
\end{figure*}
\clearpage

\clearpage
\begin{figure*}
\epsscale{1.00}
\plotone{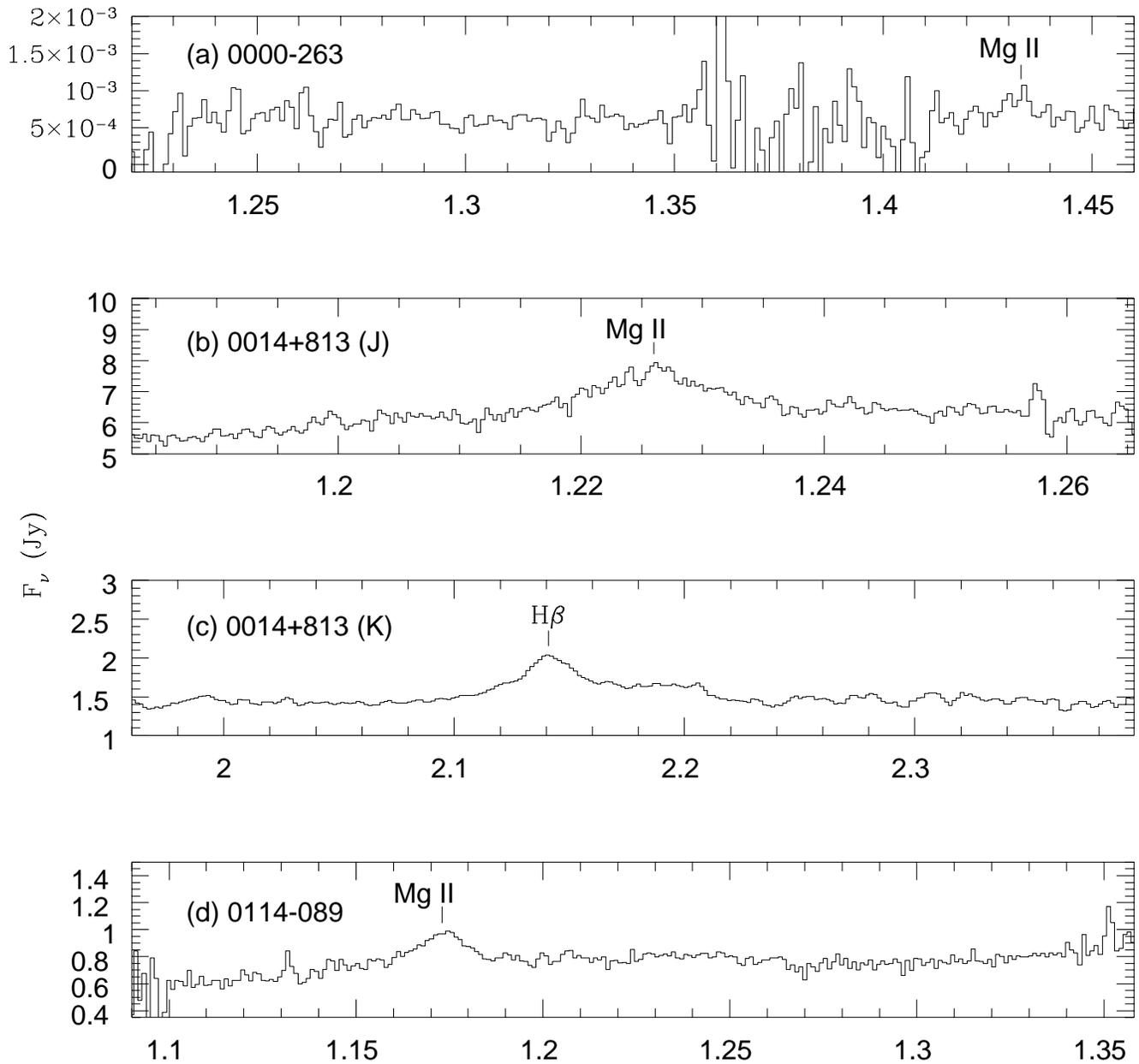}
\caption{
Infrared QSO spectra, line identifications as listed in
Table~\ref{table-z} are marked 
\label{fig:nlz} }
\end{figure*}
\clearpage

\clearpage
\begin{figure*}
\epsscale{1.00}
\plotone{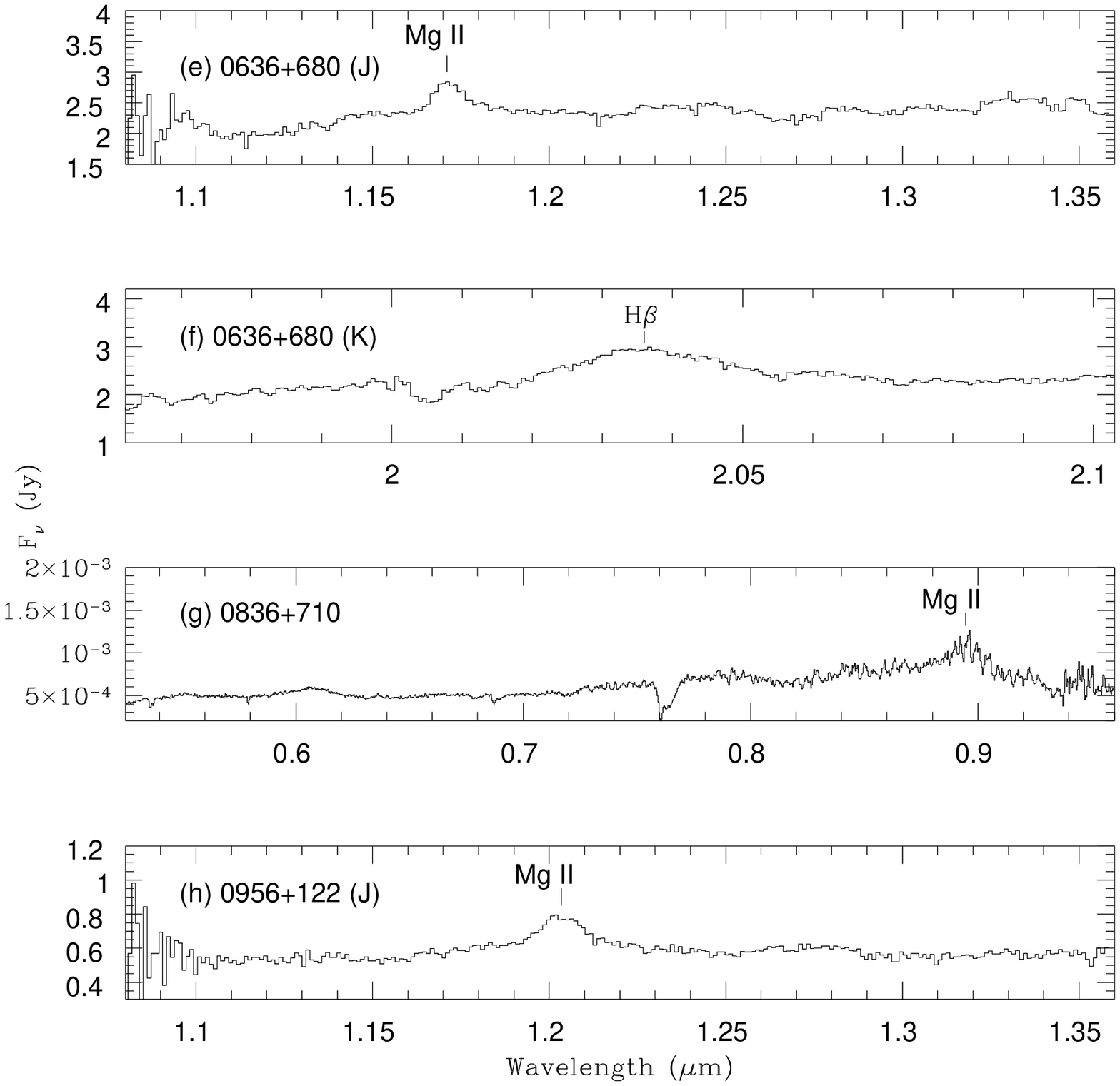}
\end{figure*}
\clearpage

\clearpage
\begin{figure*}
\epsscale{1.00}
\plotone{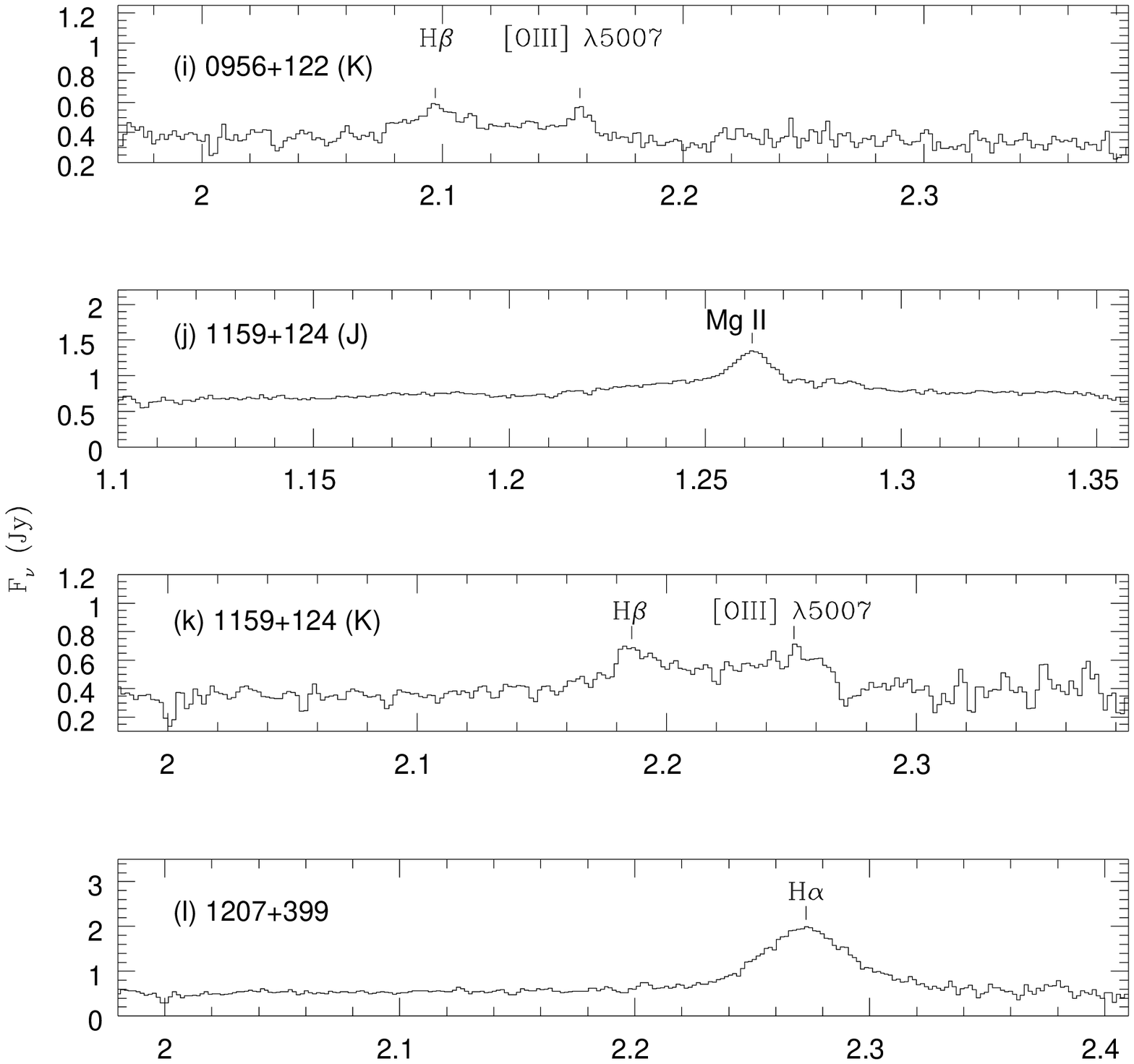}
\end{figure*}
\clearpage

\clearpage
\begin{figure*}
\epsscale{1.00}
\plotone{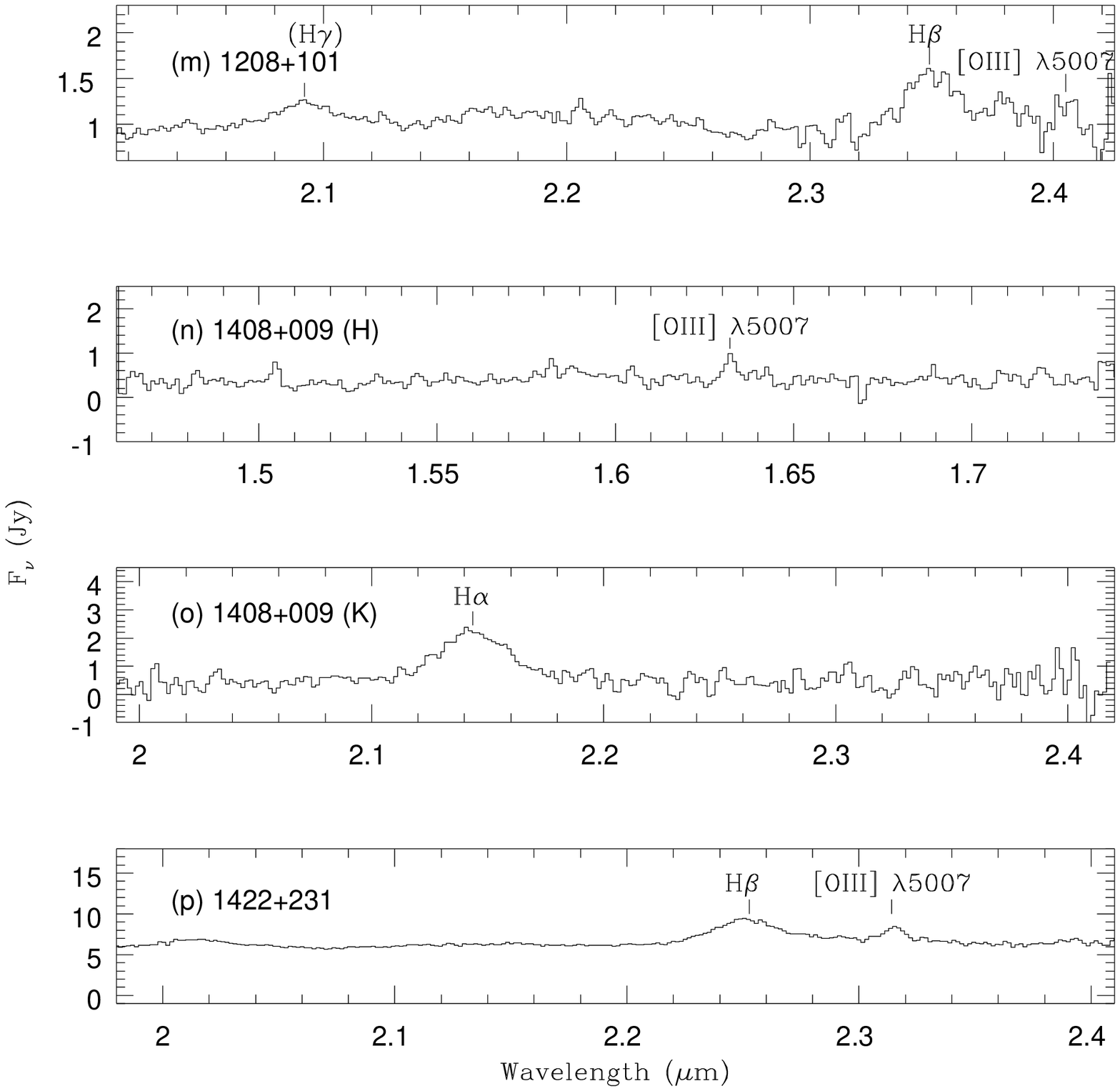}
\end{figure*}
\clearpage

\clearpage
\begin{figure*}
\epsscale{1.00}
\plotone{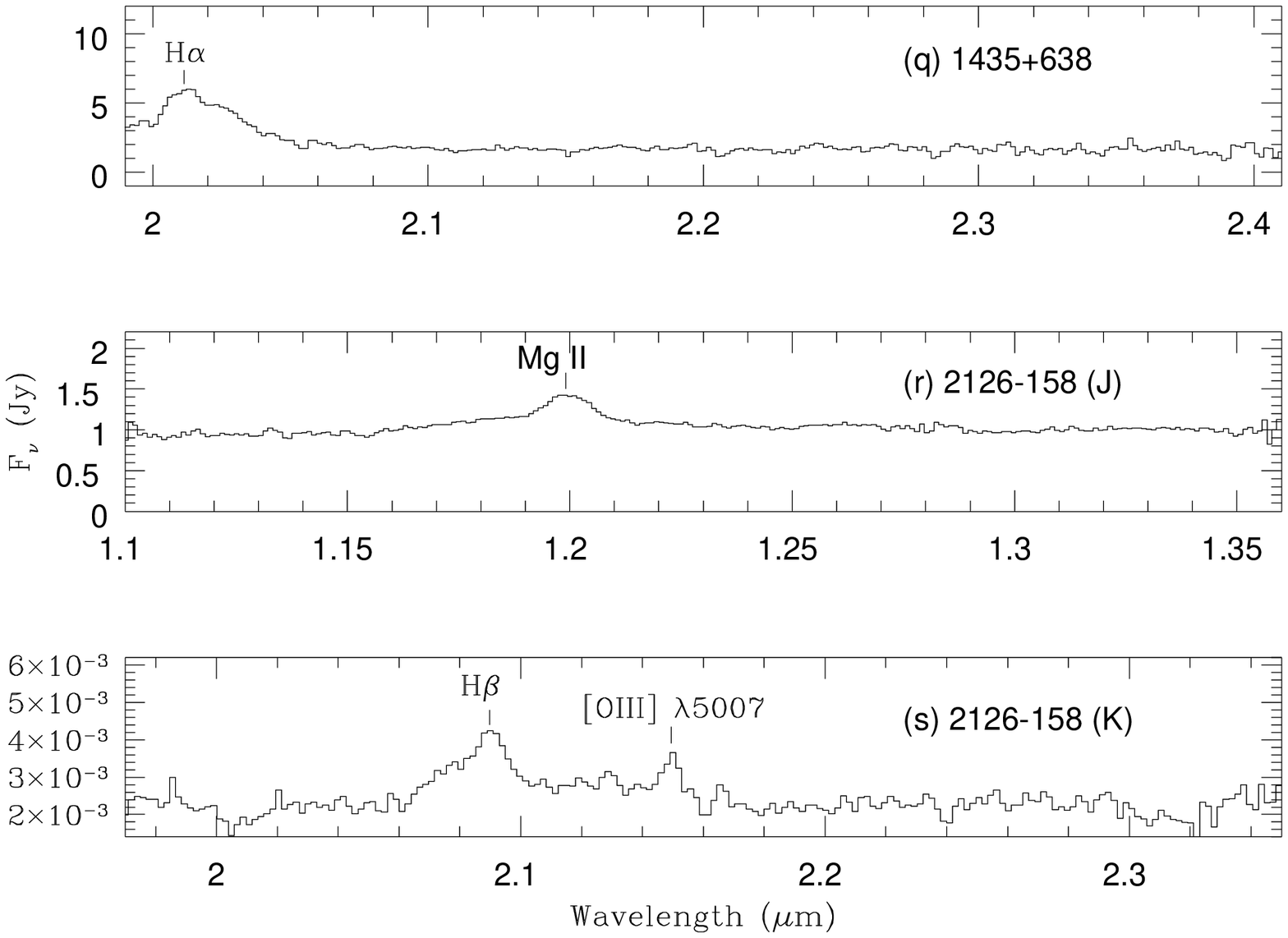}
\end{figure*}
\clearpage

\clearpage
\begin{figure*}
\epsscale{1.00}
\plotone{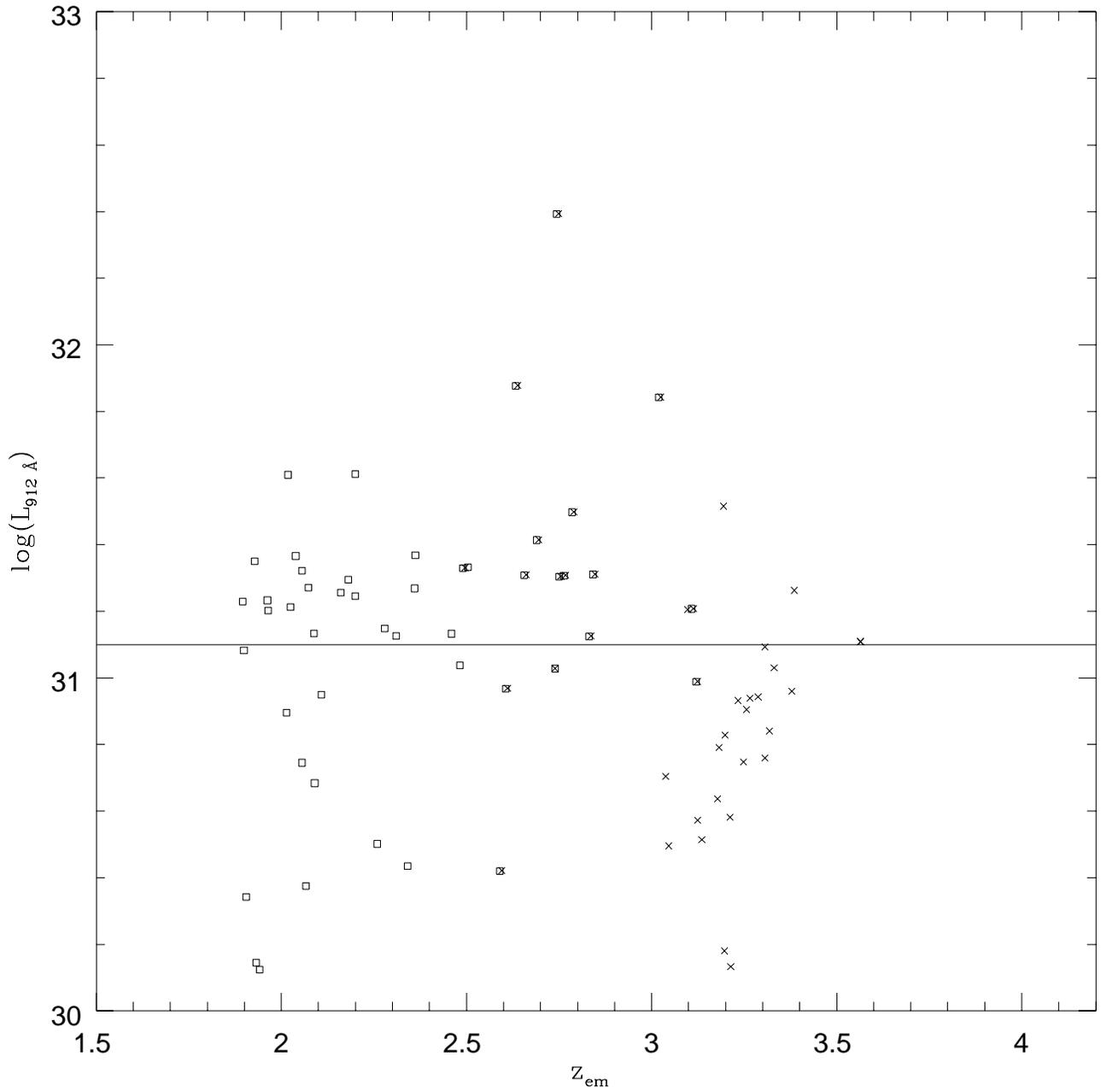}
\caption{
Lyman limit luminosity versus redshift for the proximity effect dataset;
squares- QSOs from which low redshift line sample was taken;
crosses- QSOs from which high redshift line sample was taken; the line
marks the boundary between low and high luminosity QSOs \label{fig:zlum} }
\end{figure*}
\clearpage

\clearpage
\begin{figure*}
\epsscale{1.00}
\plotone{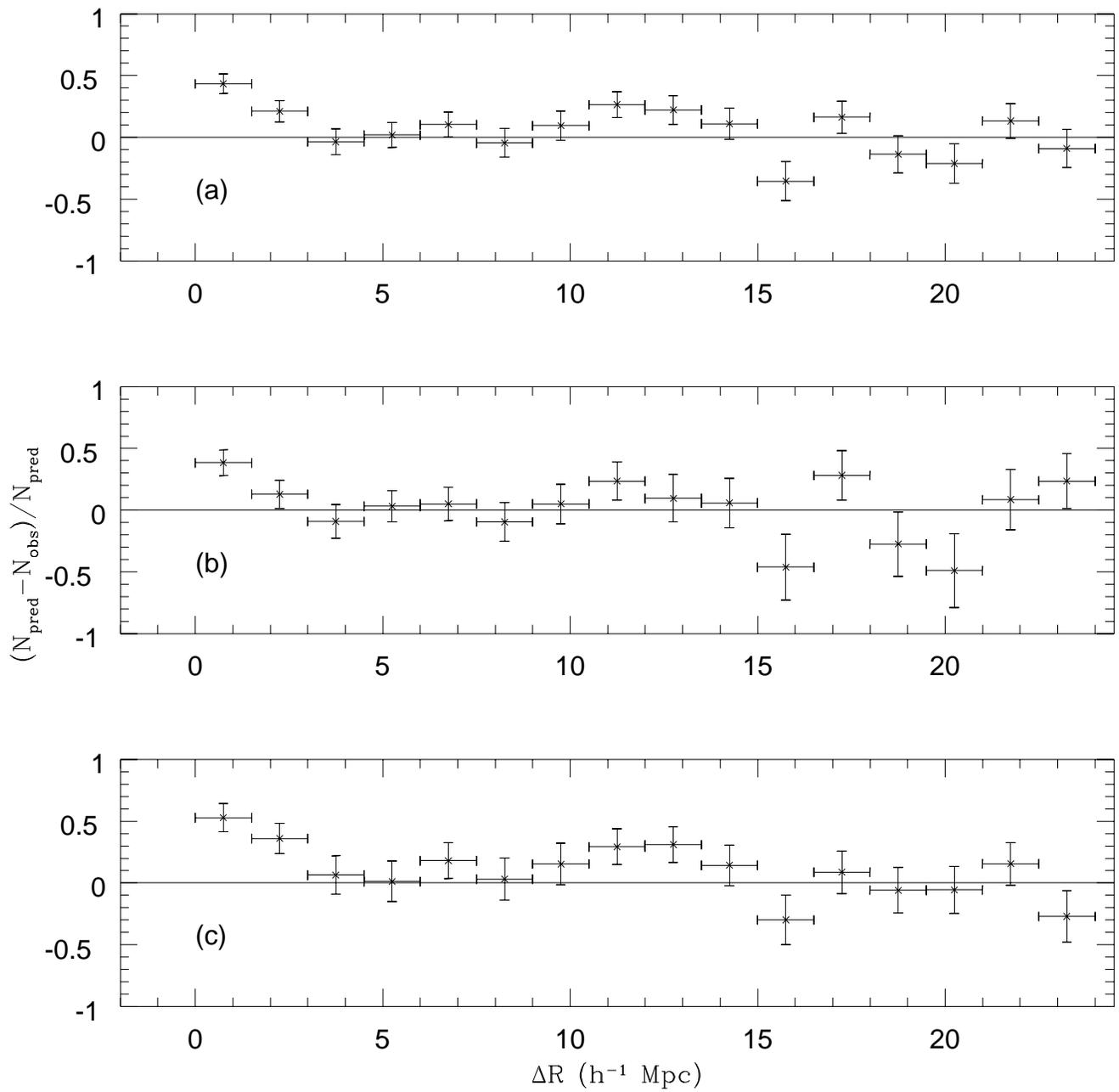}
\caption{
Relative deficit of lines with respect to the number predicted by
Equ.~\ref{eq:dndz} versus
distance from the QSO for lines with rest equivalent width greater than
0.32 $\AA$ (a) total sample; (b) low luminosity QSOs; (c) high luminosity
QSOs \label{fig:npred} }
\end{figure*}
\clearpage

\clearpage
\begin{figure*}
\epsscale{1.00}
\plotone{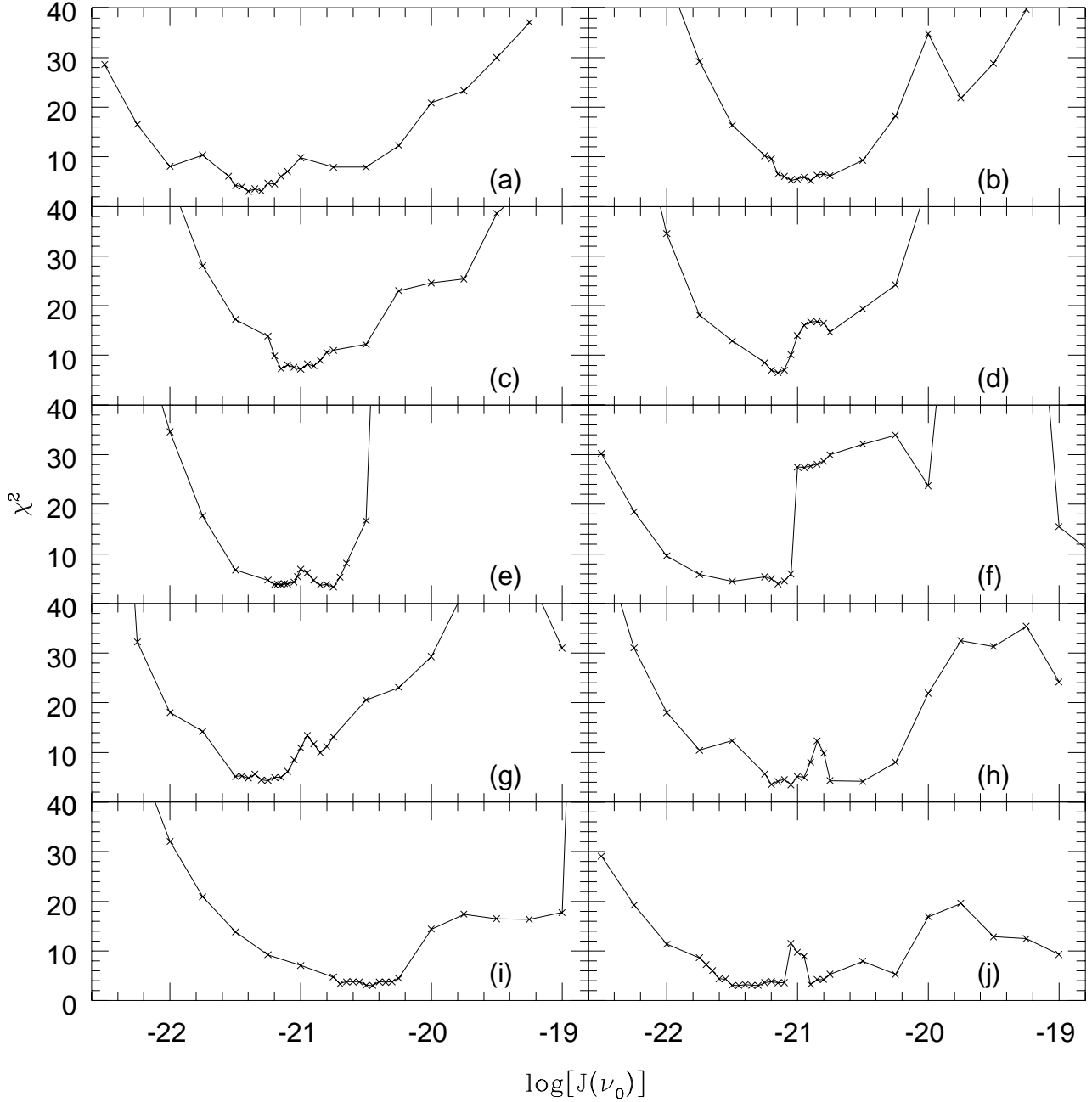}
\caption{
$\chi^{2}$ of binned data with respect to the ionization model with
a constant $J(\nu_{0})$ versus log[$J(\nu_{0})$]: (a) DB96 sample; (b) all lines,
Ly$\alpha$ QSO redshifts; (c) all lines, using narrow line redshifts
where available, Ly$\alpha$ redshifts otherwise; (d) all lines, narrow line
redshifts where available, Ly$\alpha$ redshifts + 400 km s$^{-1}$ otherwise;
(e) high z lines, QSO redshifts as in case (d); (f) low z lines,  QSO redshifts
as in case (d); (g) high luminosity QSOs, QSO redshifts as in case (d);
(h) low  luminosity QSOs, QSO redshifts as in case (d);
(i) weak lines only: 0.16 $\AA < W <$ 0.32 $\AA$, QSO redshifts as in case (d);
(j) lines from QSOs with damped Ly$\alpha$ systems only, QSO redshifts as in
case (d) \label{fig:chi2} }
\end{figure*}
\clearpage

\clearpage
\begin{figure*}
\epsscale{1.00}
\plotone{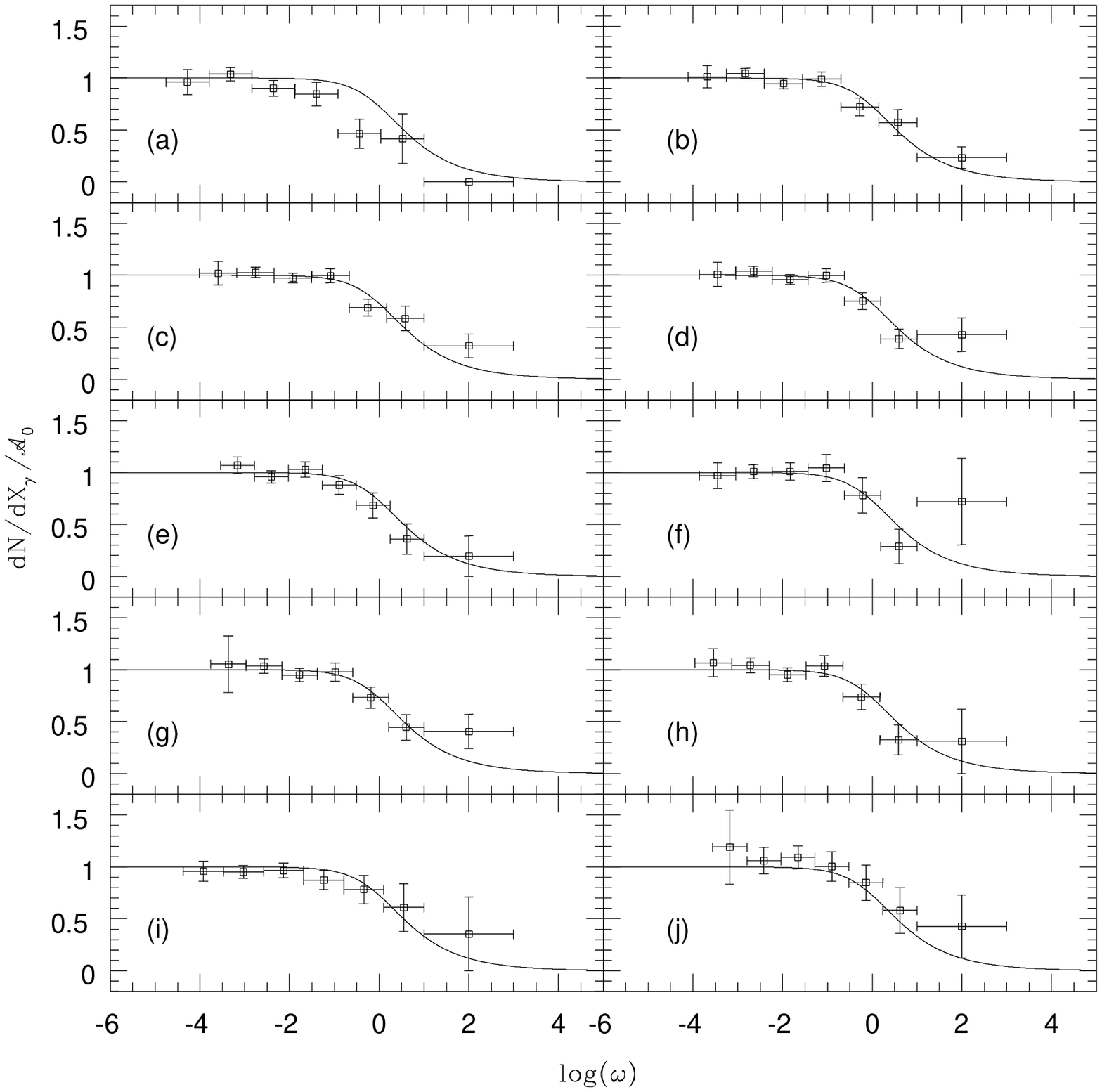}
\caption{
Number distribution per coevolving redshift coordinate for the best
fit values of $J(\nu_{0})$ (BDO method); (a-j) same as Fig.~\ref{fig:chi2}
\label{fig:dndx} }
\end{figure*}
\clearpage

\clearpage
\begin{figure*}
\epsscale{1.00}
\plotone{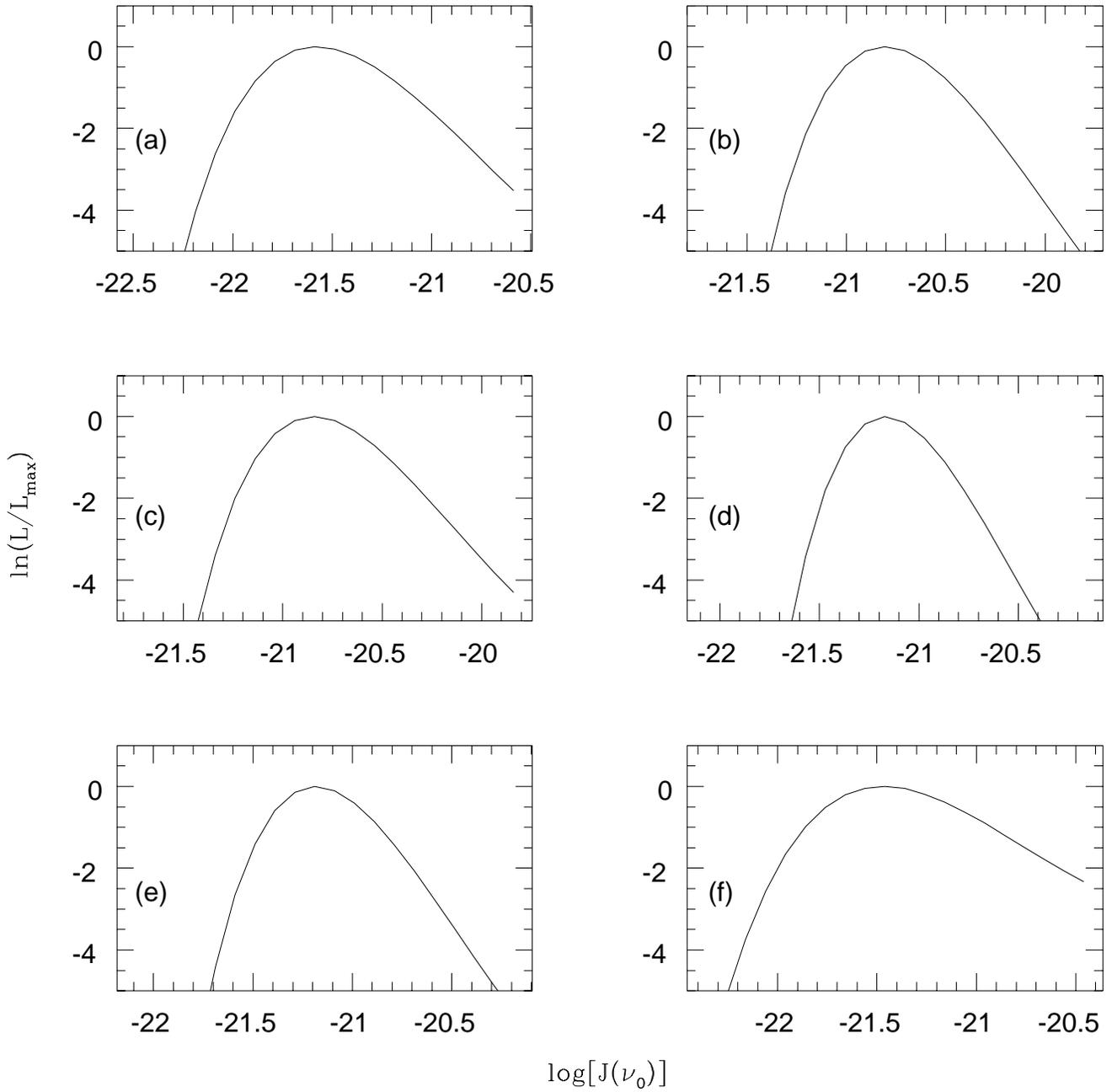}
\caption{
Likelihood function versus log[$J(\nu_{0})$]; (a-f) same as Fig.~\ref{fig:chi2}
\label{fig:like} }
\end{figure*}
\clearpage

\clearpage
\begin{figure*}
\epsscale{1.00}
\plotone{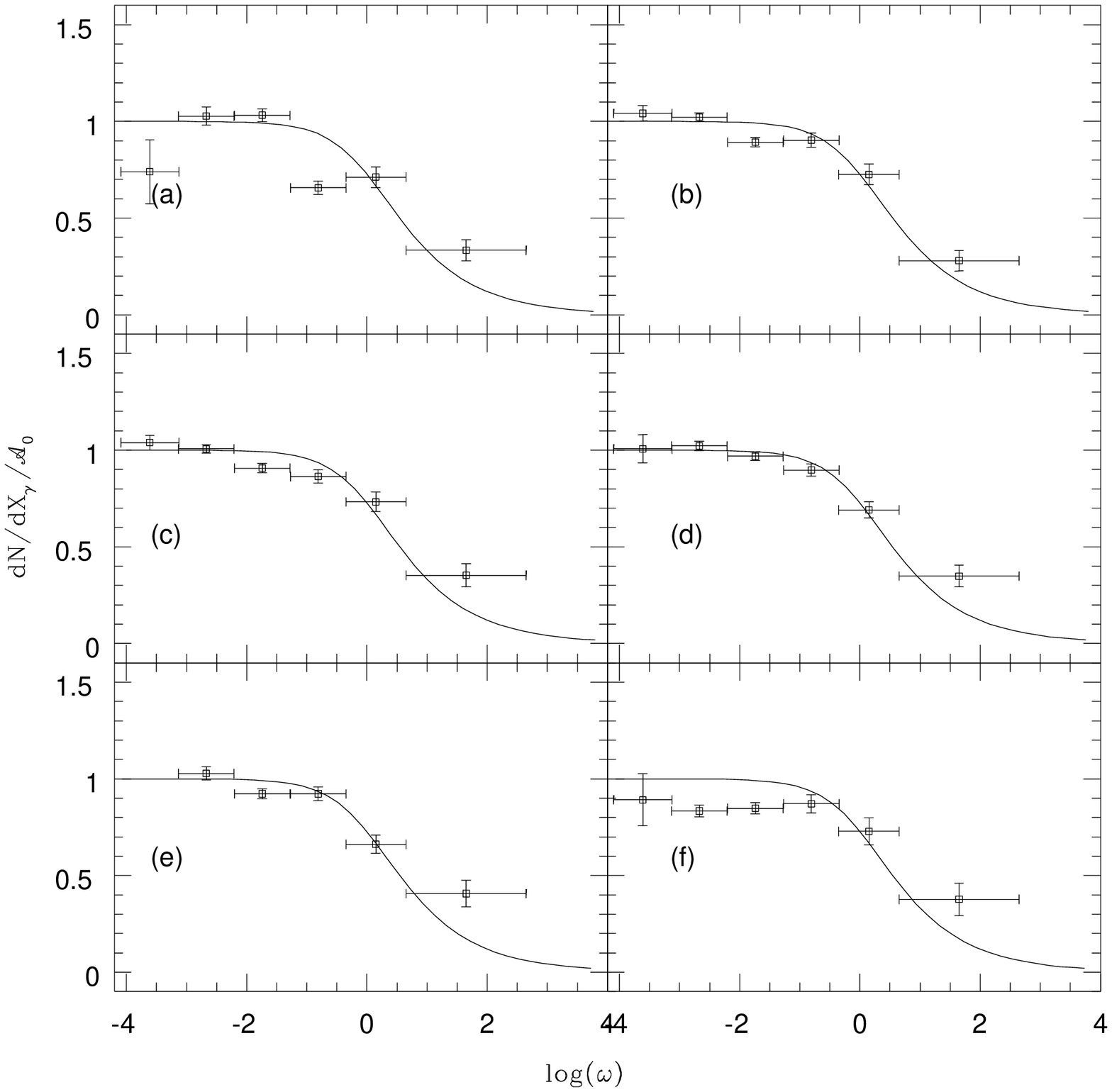}
\caption{
Number distribution per coevolving redshift coordinate for the best
fit values of $J(\nu_{0})$ (KF method); (a-f) same as Fig.~\ref{fig:chi2}
\label{fig:dndxl} }
\end{figure*}
\clearpage

\clearpage
\begin{figure*}
\epsscale{1.00}
\plotone{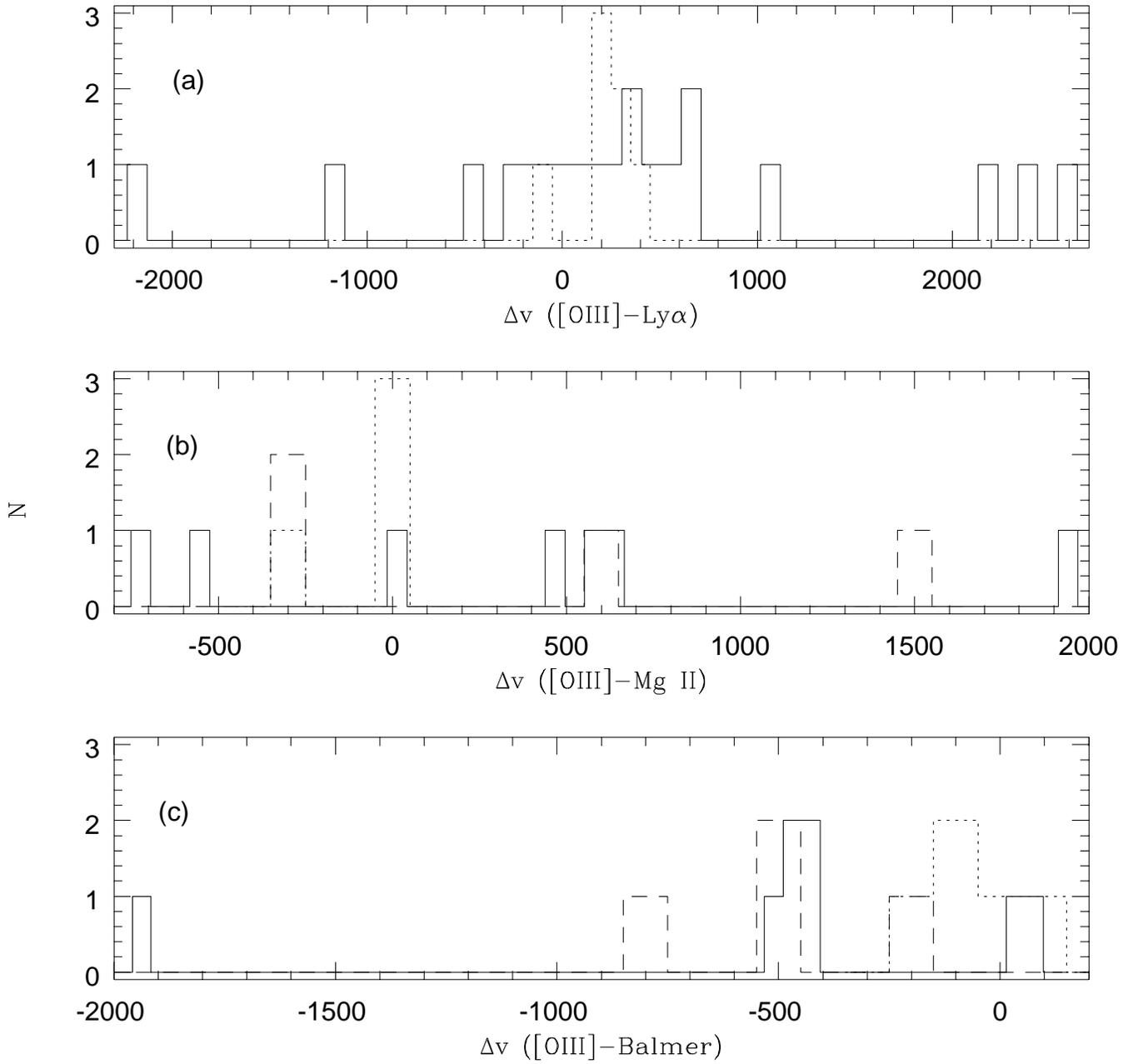}
\caption{
Histograms of redshift differences with respect to
the [OIII] $\lambda$5007 line (a) Ly$\alpha$; (b) Mg II; (c) Balmer lines ;
dotted lines- data of Laor et al. (1995), dashed lines- data
of Nishihara et al. (1997) \label{fig:zhist} }
\end{figure*}
\clearpage

\clearpage
\begin{figure*}
\epsscale{1.00}
\plotone{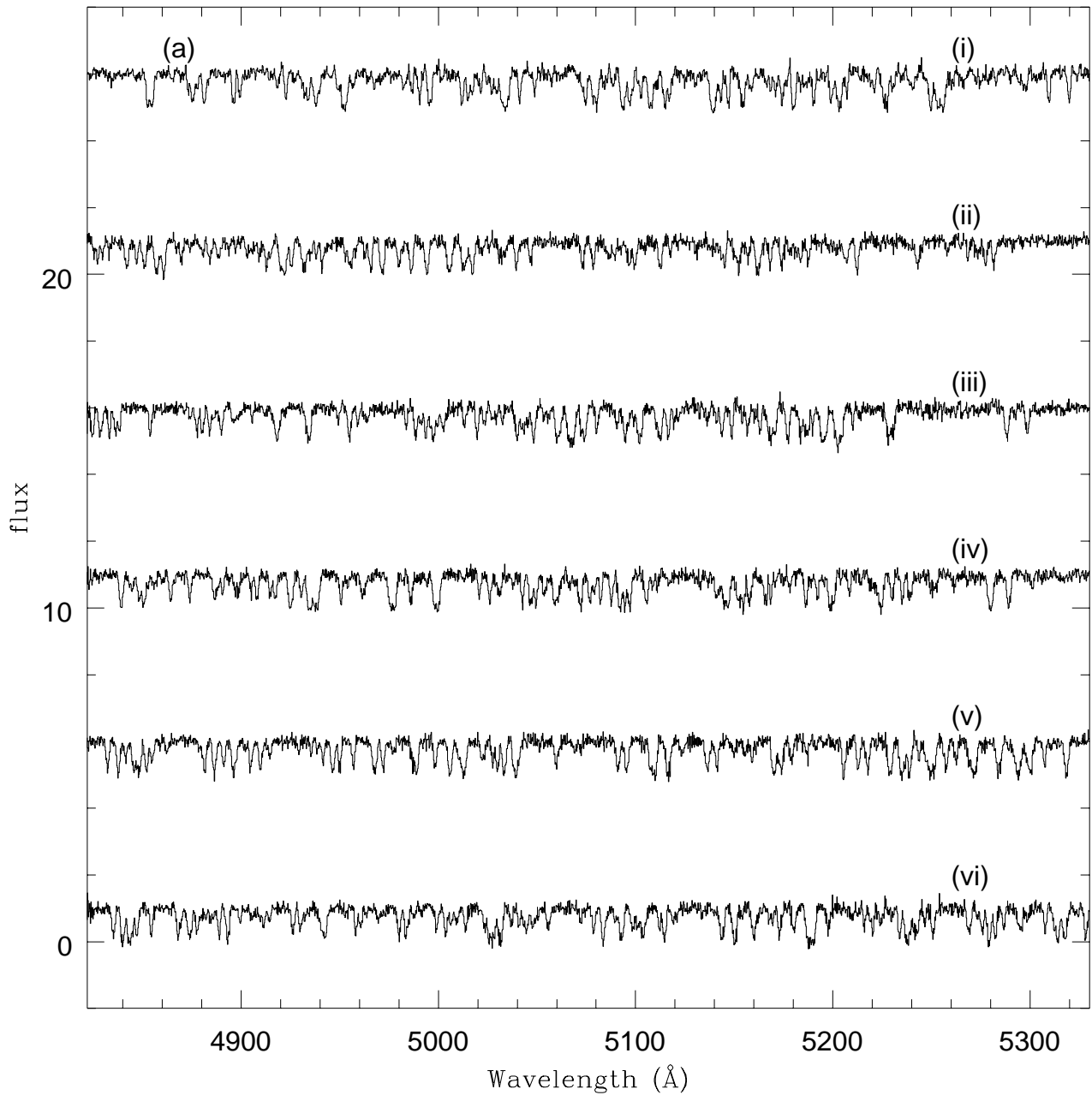}
\caption{
Sample simulation spectra plotted with data, flux
scale is arbitrary. (a)0014+813:
(i)data, (ii)input log[$J(\nu_{0})$]=-23.0, (iii)input log[$J(\nu_{0})$]=-22,
(iv)input log[$J(\nu_{0})$]=-21.3, (v)input log[$J(\nu_{0})$]=-20.0,
(vi)input log[$J(\nu_{0})$]=-19.0;  (b)1700+643: (i)-(vi) same as in (a) 
\label{fig:simspec} }
\end{figure*}
\clearpage

\clearpage
\begin{figure*}
\epsscale{1.00}
\plotone{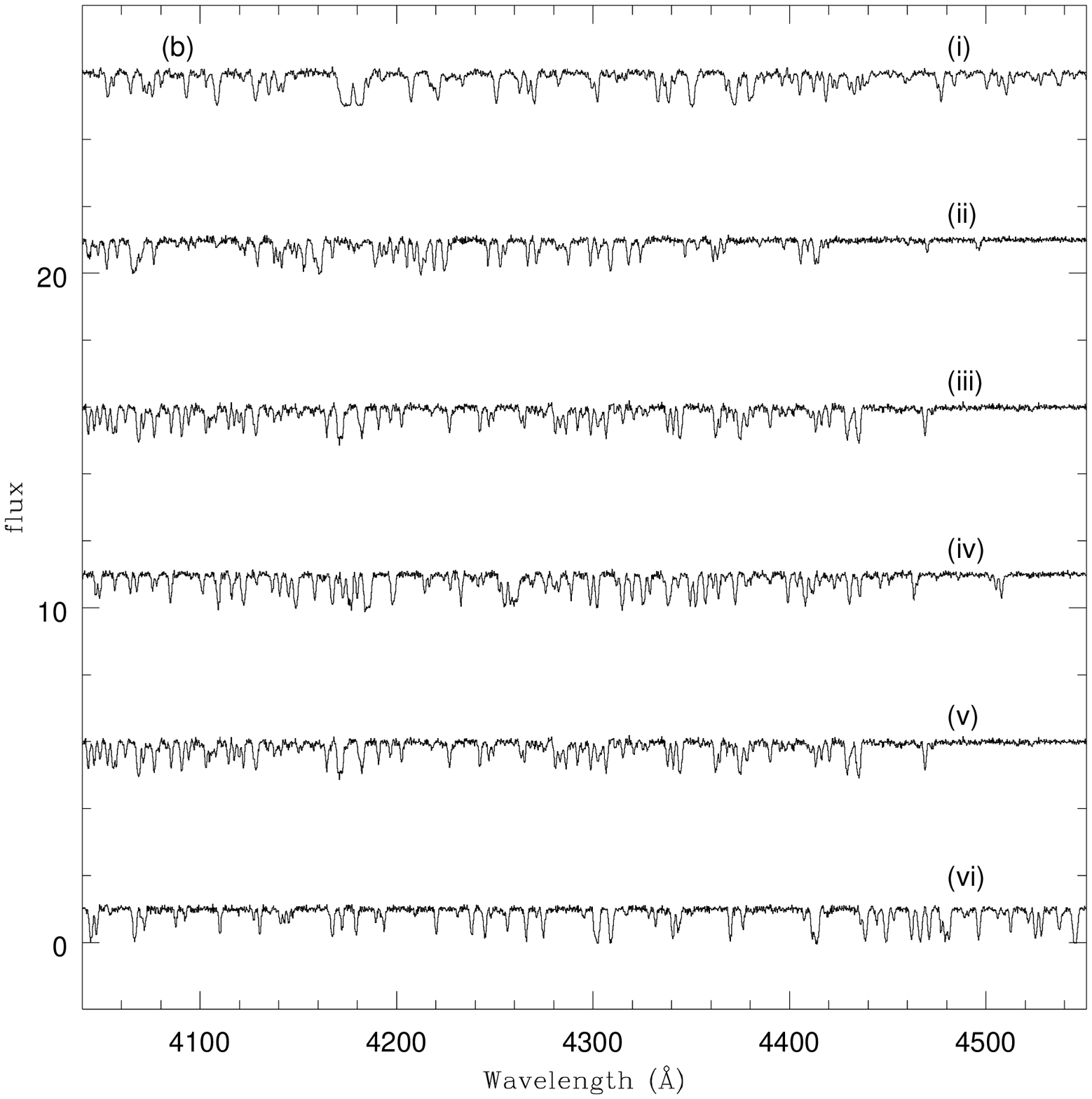}
\end{figure*}
\clearpage

\clearpage
\begin{figure*}
\epsscale{1.00}
\plotone{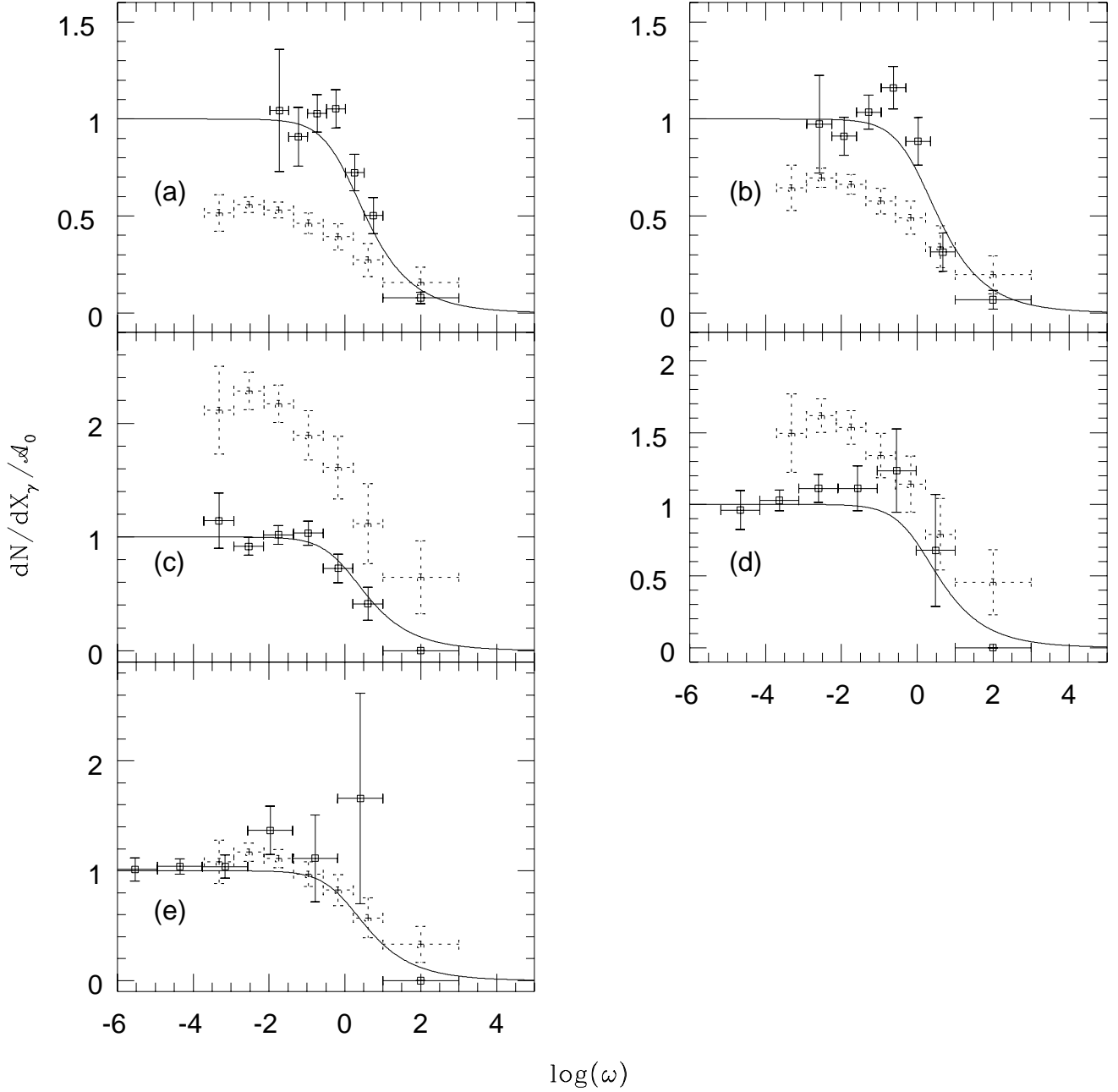}
\caption{
Number distribution per coevolving redshift coordinate for the best
fit values of $J(\nu_{0})$ listed in Table~\ref{table-sim};
solid lines- simulation, dotted lines- data, scaled by the
relevant value of ${\cal A}_{0}$ in Table~\ref{table-sim}:
(a)input log[$J(\nu_{0})$]=-23.0; (b)input log[$J(\nu_{0})$]=-22;
(c)input log[$J(\nu_{0})$]=-21.3; (d)input log[$J(\nu_{0})$]=-20.0;
(e)input log[$J(\nu_{0})$]=-19.0 \label{fig:simdndx} }
\end{figure*}
\clearpage

\clearpage
\begin{figure*}
\epsscale{1.00}
\plotone{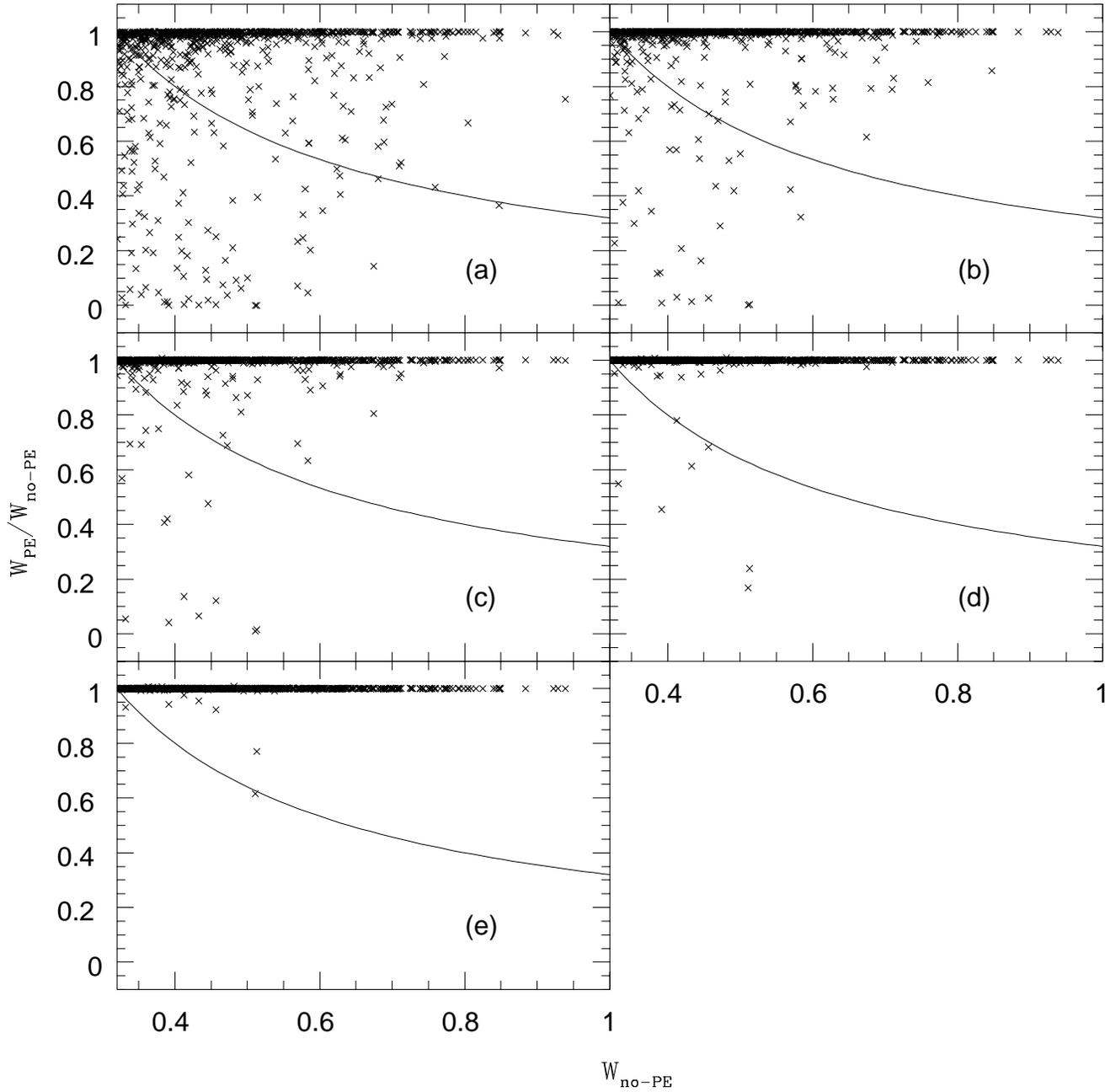}
\caption{
Curve of growth effects: ratio of proximity effect to nonproximity effect
rest equivalent width for all lines versus nonproximity effect rest equivalent
width; solid line represents the detection threshold $W_{PE}$= 0.32 $\AA$;
(a)-(e) as in Fig.~\ref{fig:simdndx} \label{fig:cog} }
\end{figure*}
\clearpage

\clearpage
\begin{figure*}
\epsscale{1.00}
\plotone{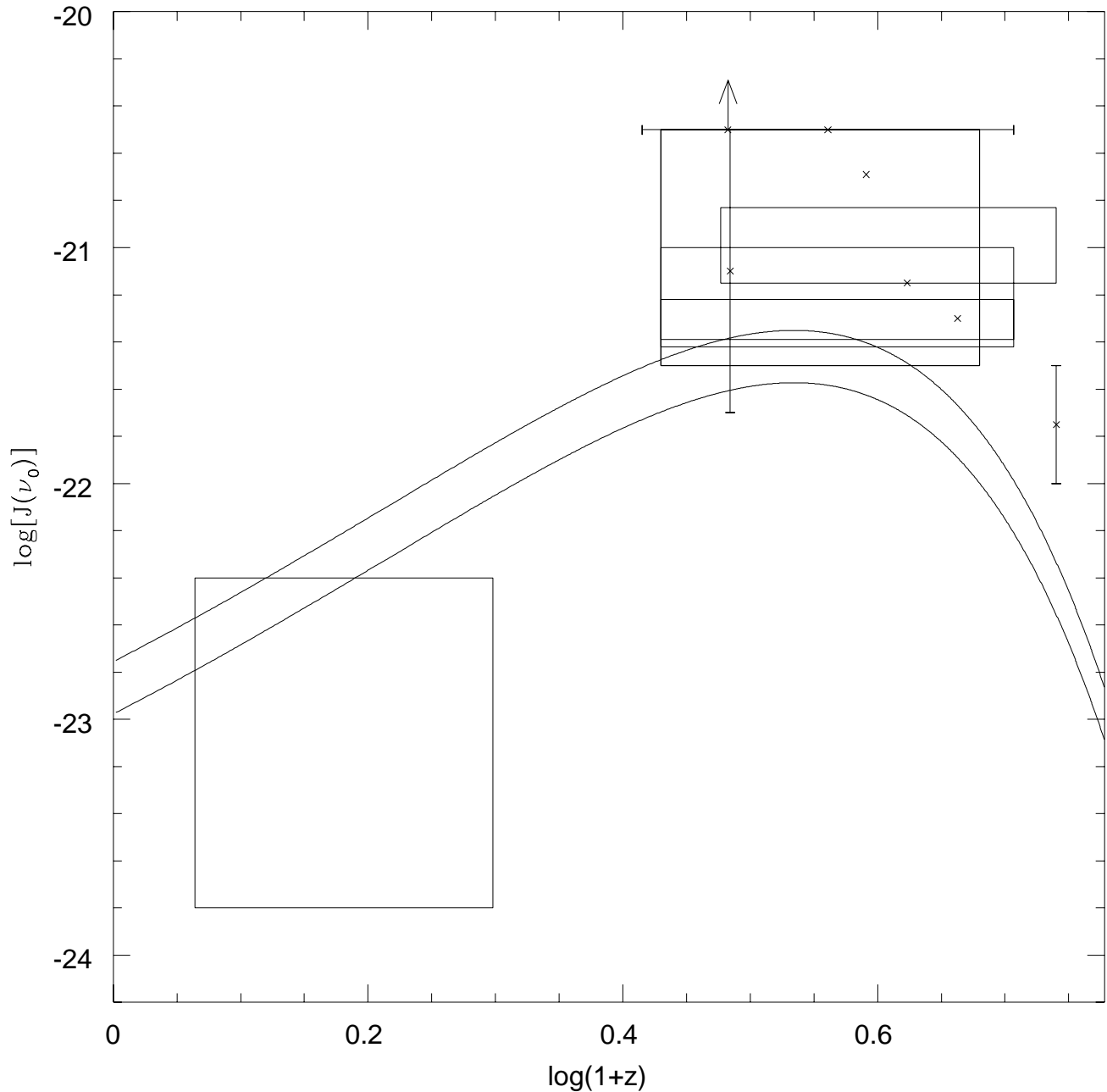}
\caption{
Measurements of log[$J(\nu_{0})$] vs. redshift:  points and error bars
are taken from
Table~\ref{table-lit}. The
upper limit set by 
Bunker et al. (1998) at z$\sim$3 is included. 
Measurements over extended redshift ranges and the errors in those
measurements are indicated by boxes.
The solid curves are
derived from the Haardt \& Madau (1996) model for the HI 
photoionization rate as a
function of redshift for QSO spectral indicies of 0 (lower curve)
and 2 (upper curve).  Overall, measurements at z=2-3 agree well
with one another and with the predictions of the Haardt \& Madau (1996) model.
\label{fig:sum}}
\end{figure*}
\clearpage

\clearpage
\begin{figure*}
\epsscale{1.00}
\plotone{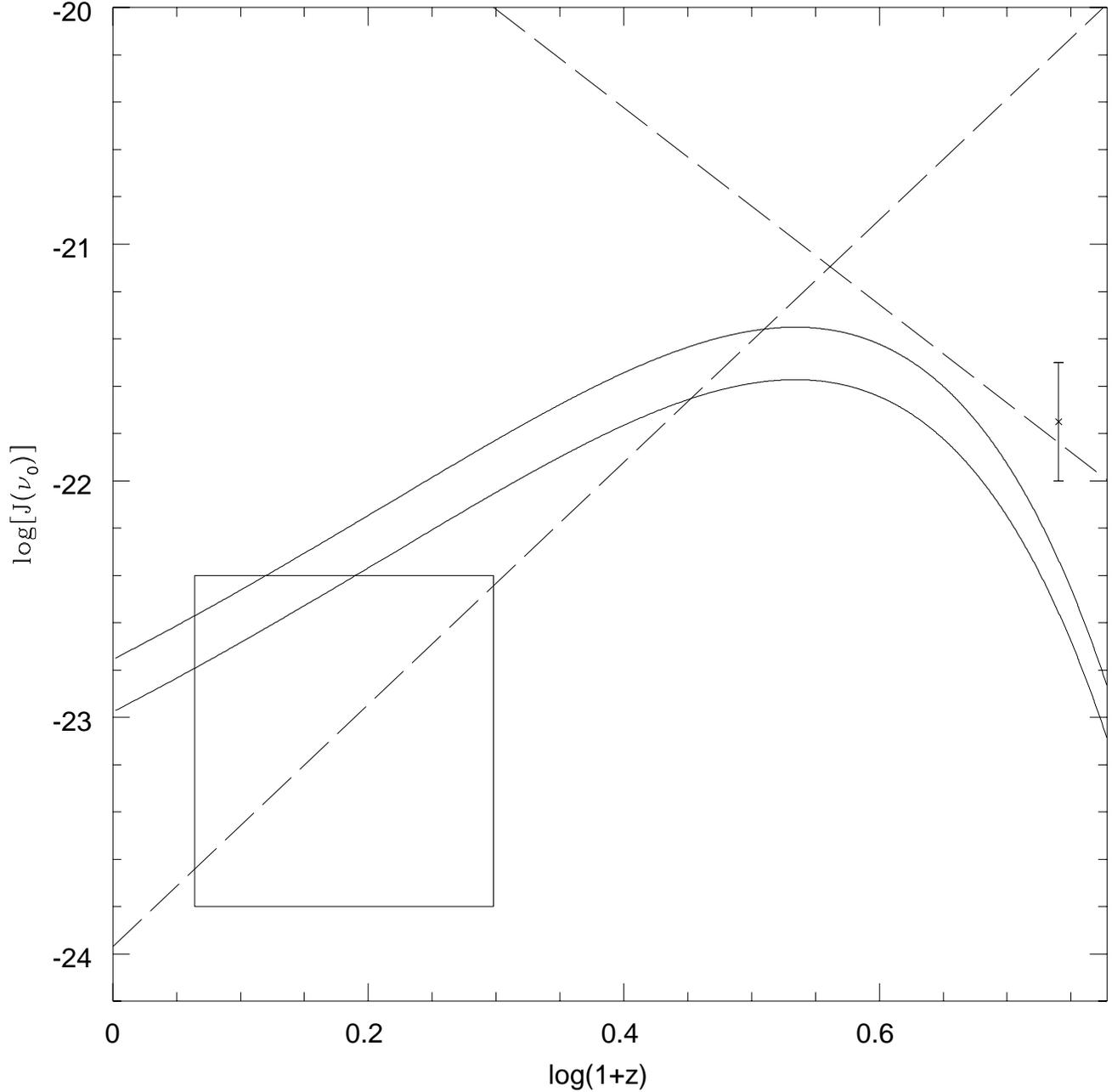}
\caption{Power law fits to log[$J(\nu_{0})$] as a function of redshift: 
$J(\nu_{0},z)=J(\nu_{0},0)(1+z)^{j}$.
The dashed lines indicate the two lowest 
$\chi^{2}$ fits to the data: 
($j$,log[$J(\nu_{0},0)$])= (5.12,-23.97) and (-4.16,-18.76).
The solid curves are the Haardt \& Madau (1996) models as shown in Figure~\ref{fig:sum}.
The Haardt \& Madau (1996)
models are turning over at the redshift of the data, precluding a strong constraint on 
the parameters $j$ and $J(\nu_{0},0)$; but the lowest $\chi^{2}$ fit extends to low redshift
to match the Kulkarni \& Fall (1993) measurement shown by the box, while the next
lowest $\chi^{2}$ fit extends to high redshift to match the Williger et al. (1994) measurement, the
point at z=4.5.
\label{fig:pl}}
\end{figure*}
\clearpage

\pagebreak

\begin{deluxetable}{llccc}
\tablefontsize{\footnotesize}
\tablecolumns{5}
\tablewidth{40pc}
\tablecaption{Spectrophotometry Observations \label{table-specphotobs}}
\tablehead{
\colhead{QSO} &  
\colhead{Date} & \colhead{Exposure} & \colhead{Airmass} &
\colhead{Wavelength Coverage}  \\
\colhead{} & 
\colhead{} & \colhead{(seconds)} & \colhead{} & \colhead{($\AA$)}  }
\startdata
 0006+020  &29Nov1994 &1800 &1.15 &3150-6385 \nl
 0027+018  &22Sep1992 &1800 &1.28 &3467-6475 \nl
 0037-018  &29Nov1994 &2400 &1.27 &3150-6385 \nl
 0049+007  &29Nov1994 &1800 &2.05 &3125-6380 \nl
 0123+257  &29Nov1994 &1800 &1.55 &3125-6380 \nl 
 0153+744  &29Nov1994 &1800 &1.38 &3125-6380  \nl
 0348+061  &22Sep1992 &1800 &1.13 &3465-6475  \nl 
 1323-107  &28Mar1995 &1800 &1.56 &3115-6400  \nl 
 1346-036  &28Mar1995 &1800 &1.27 &3115-6400  \nl 
 1422+231  &22Apr1996 &1800 &1.31 &5235-7554  \nl
 1603+383\tablenotemark{a} &04July1995 &450  &1.03 &3663-7544 \nl 
 2134+004  &22Sep1992 &1800 &1.29 &3465-6483  \nl
 2251+244  &29Nov1994 &1800 &1.01 &3150-6385  \nl 
 2254+022  &22Sep1992 &1800 &1.18 &3470-6480  \nl 
\tablenotetext{}
{\footnotesize{Instrument Set-up for 
1422+231:  SO B\&C, 600 l mm$^{-1}$ 1$^{st}$ order, 
$\lambda_{b}$=6681 $\AA$, 1.5$\arcsec$ slit;  
for 1603+383: 
FLWO FAST, 300 l mm$^{-1}$ 1$^{st}$ order, $\lambda_{b}$=4750,
3$\arcsec$ slit}}
\tablenotetext{\footnotesize{a}}{\footnotesize{spectrum donated by Hamburg/CfA Bright Quasar Survey
in advance of publication}}
\enddata
\end{deluxetable}

\begin{deluxetable}{llcccc}
\tablefontsize{\footnotesize}
\tablecolumns{6}
\tablewidth{40pc}
\tablecaption{Summary of Narrow Emission Line Observations
\label{table-z2obs}}
\tablehead{
\colhead{Name}  & \colhead{V}&
\colhead{Instrument}   & \colhead{Date}  & \colhead{Exposure (sec.)} &
\colhead{Wavelength Coverage ($\mu$m)} }
\startdata
0000-263 &17.5 &OSIRIS &27Jul1994 &4800 &1.20  - 1.46 \nl
0014+813 &16.5 &CRSP    &07Dec1993  &1200  &1.18  - 1.26  \nl
         &     &FSpec   &26Nov1993  &5280  &1.96  - 2.39   \nl 
0114-089 &17.4 &CRSP    &04Dec1993  &3180  &1.10  - 1.35   \nl
0636+680 &19.0 &CRSP    &04Dec1993  &4800  &1.09  - 1.35   \nl
         &     &CRSP    &05Dec1993  &2820  &1.96  - 2.10   \nl
0836+710 &16.5 &B\&C   &29Mar1995 &1800 &0.52 - 0.96   \nl
0956+122 &17.5 &CRSP  &04Dec1993 &8220 &1.10  -  1.35  \nl
         &     &FSpec &27Nov1993 &5280 &1.96  - 2.38  \nl 
1159+124 &17.5 &CRSP  &05Dec1993 &3180 &1.09  - 1.35  \nl 
         &     &FSpec &29Nov1993 &4320 &1.97  - 2.38  \nl 
1207+399 &17.5 &FSpec  &21May1994 &600  &1.98  - 2.41  \nl 
1208+101 &17.5 &CRSP  &06Dec1993 &4800 &2.00  - 2.42  \nl
1408+009 &18.0 &FSpec &02Apr1996 &3840 &1.46  - 1.73   \nl
         &      &     &02Apr1996 &1920 &1.99  - 2.40   \nl
1422+231 &16.5 &FSpec &21May1994 &1920 &1.98  - 2.41   \nl
1435+638 &15.0 &FSpec &02Apr1996 &1920 &1.99  - 2.40   \nl
2126-158 &17.3 &CRSP &05Dec1993 &3180 &1.08  - 1.35   \nl
         &     &OSIRIS&24Sep1994&7680 &1.96  - 2.35  \nl
\enddata 
\end{deluxetable}

\begin{deluxetable}{lcccccl}
\tablefontsize{\footnotesize}
\tablecolumns{7}
\tablewidth{42pc}
\tablecaption{Spectrophotometric Properties \label{table-specphotfinal}}
\tablehead{
\colhead{QSO}  &\colhead{N$_{HI}$ (10$^{20}$ cm$^{-2}$)}  
&\colhead{f$_{\nu}^{obs}$(912 $\AA$)} 
&\colhead{$\alpha$}   &\colhead{f$_{\nu}^{obs}$}  
&\colhead{f$_{\nu}$(912 $\AA$)} 
&\colhead{Ref.} \\
\colhead{(a)}  &\colhead{(b)}  &\colhead{(c)} &
\colhead{(d)}   & \colhead{(e)}  & \colhead{(f)} &
\colhead{(g)} }
\startdata
0006+020*&3.02  &    &0.26  &354 (1450 $\AA$) &313    &1  \nl
0027+014 &2.93  &    &-0.38  &219 (1450 $\AA$)&183    &1  \nl
0037-018 &2.81  &    &-0.27 &45  (1450 $\AA$) &51     &1  \nl
0049+007 &2.67  &    &0.31  &324 (1450 $\AA$) &280    &1  \nl
0123+257*&6.88  &    &1.12  &237 (1450 $\AA$) &141    &1\nl
0150-202*&1.29  &    &      &529 (1430 $\AA$) &430 &2,3\nl
0153+744*&22.74 &    &0.18  &1023 (1450 $\AA$)&940 &1\nl
0226-038 &2.35  &    &      &582 (1800 $\AA$) &425 &4\nl
0348+061 &12.33 &    &0.12  &513 (1450 $\AA$) &485 &1\nl
0400+258 &7.82  &    &1.54  &                 &199 &5\nl
0747+610 &4.77  &    &      &500 (1800 $\AA$) &365 &4\nl
0819-032 &6.16  &    &0.33  &63  (1450 $\AA$) &54  &6\nl
0836+710*&2.93  &    &      &                 &652 &  \nl
0848+155 &3.14  &    &0.07  &198 (1450 $\AA$) &191 &7,8 \nl
0936+368 &1.36  &    &      &                 &386 &    \nl
0952+335 &1.37  &    &      &                 &370 &     \nl
0955+472*&1.04  &    &      &                 &188 &    \nl
0956+122 &3.10  &140 &0.49  &448 (1450 $\AA$) &356 &9   \nl
1009+299 &2.30  &    &      &                 &1217&      \nl
1207+399 &2.10  &    &0.59  &319 (1450 $\AA$) &242 &1,8  \nl
1210+175*&2.67  &    &      &                 &285 &   \nl
1231+294 &1.54  &    &      &                 &980 &     \nl
1323-107 &2.64  &    &-0.30 &303 (1450 $\AA$) &349 &1    \nl
1329+412*&0.99  &    &0.33  &                 &750 &10\nl
1337+285*&1.17  &    &      &                 &339 &     \nl
1346-036 &2.51  &    &0.091 &458 (1450 $\AA$) &439 &1    \nl 
1358+115*&1.81  &    &1.10  &345 (1450 $\AA$) &207 &6    \nl
1406+492 &1.77  &    &      &                 &392 &     \nl
1408+009 &3.04  &    &0.91  &99  (1450 $\AA$) &64  &1 \nl
1421+330 &1.23  &58  &0.54  &914 (1450 $\AA$) &711 &11,7\nl
1422+231*&2.52  &    &-1.21 &211 (1450 $\AA$) &371 &1   \nl
1435+638 &1.68  &55  &      &1244 (1800 $\AA$)&909 &12,4\nl
1603+383*&1.32  &    &0.36  &550 (1450 $\AA$) &464 &1,13    \nl
1604+290 &3.24  &    &      &                 &428 &    \nl
1715+535 &2.69  &36  &1.26  &875 (1800 $\AA$) &371 &11,10,4\nl
2134+004 &4.03  &    &0.04  &35  (1450 $\AA$) &34  &1,14     \nl
2251+244*&5.18  &    &1.53  &243 (1450 $\AA$) &119 &1        \nl
2254+024 &5.32  &    &0.20  &116 (1450 $\AA$) &106 &1 \nl
2310+385 &10.62 &    &      &                 &419 &   \nl
2320+079 &5.04  &    &      &                 &306 &   \nl
2329-020 &4.45  &    &      &                 &451 &    \nl
\tablenotetext{\footnotesize{a}}{\footnotesize{QSO name; an asterisk denotes a metal line system within
5000 km s$^{-1}$ of the QSO emission redshift; in the case of
1422+231, the QSO is a known lens}}
\tablenotetext{\footnotesize{b}}{\footnotesize{Galactic N$_{HI}$ in units of $10^{20}$ cm$^{-2}$ from
program COLDEN using Stark et al. (1992)}}
\tablenotetext{\footnotesize{c}}{\footnotesize{Observed flux in $\mu$Jy at the Lyman limit from reference 
in (g)}}
\tablenotetext{\footnotesize{d}}{\footnotesize{Observed spectral index between Ly$\alpha$ and C IV
emission lines or in the vicinity of the flux listed in (e) from reference
in (g);  in general, values are based upon spectra corrected for
Galactic reddening if E(B-V) $\gtrsim$ 0.03}} 
\tablenotetext{\footnotesize{e}}{\footnotesize{Observed flux in $\mu$Jy at the rest wavelength indicated in 
parentheses from reference in (g)}}
\tablenotetext{\footnotesize{f}}{\footnotesize{Extrapolated Lyman limit flux in $\mu$Jy from measured flux 
in (e), when available, or V magnitude given in Table 1 of Paper I.;
if no observed spectral index available, value of 0.46 used (Francis 1996)} }
\tablenotetext{\footnotesize{g}}{\footnotesize{REFERENCES: (1)this paper;
(2)MacAlpine \& Feldman 1982;
(3)Griffith et al. 1994; (4) Steidel \& Sargent 1991;
(5)Cheng,Gaskell, \& Koratkar 1991;
(6)Pei, Fall \& Bechtold 1991; (7)Uomoto 1984; (8)Barthel et al. 1988;
(9) Sargent, Steidel, \& Boksenberg 1989;
(10)Baldwin, Wampler, \& Gaskell 1989;
(11)Koratkar, Kinney, \& Bohlin 1992; (12)Lanzetta, Turnshek \& Sandoval 1993
(13) Hamburg/CfA Bright Quasar Survey (unpublished);
(14)Perez, Penston, \& Moles 1989}}
\enddata
\end{deluxetable}

\begin{deluxetable}{clclccccc}
\tablefontsize{\footnotesize}
\tablecolumns{9}
\tablewidth{0pc}
\tablecaption{Measurements of $J(\nu_{0})$ 
\label{table-jnu} } 
\tablehead{\colhead{Sample} &\colhead{${\cal N}_{lines}$} & 
\colhead{$\gamma$,norm.}  &
\colhead{method} & \colhead{log[$(J(\nu_{0})$]}  & \colhead{$\chi^{2}$}  &
\colhead{${\cal N}_{points}$} &\colhead{Q$_{\chi^{2}}$} &
\colhead{Figure} \\
\colhead{(a)}  &\colhead{(b)} &\colhead{(c)} &\colhead{(d)} &\colhead{(e)} &
\colhead{(f)} &\colhead{(g)} &\colhead{(h)} &
\colhead{(i)} }
\startdata
1 &518  &1.9260,5.8882 &BDO &-21.40$^{+1.1}_{-0.69}$  &3.05 &7 &0.88  &6(a) \nl
1 &518  &1.9260,3.9709 &ML  &-21.58$^{+0.30}_{-0.23}$ &20.3 &6 &0.0024 &8(a)\nl 
2 &1286 &1.6749,7.5723 &BDO &-20.90$^{+0.61}_{-0.48}$ &5.22 &7 &0.63  &6(b) \nl
2 &1286 &1.6749,4.6637 &ML  &-20.83$^{+0.23}_{-0.20}$ &6.32 &6 &0.38  &8(b)\nl
3 &1286 &1.6749,7.5723 &BDO &-21.00$^{+0.57}_{-0.36}$ &7.19 &7 &0.40  &6(c) \nl
3 &1286 &1.6749,4.6709 &ML  &-20.83$^{+0.24}_{-0.22}$ &7.41 &6 &0.28  &8(c) \nl 
4 &1286 &1.6749,7.5723 &BDO &-21.15$^{+0.17}_{-0.43}$ &6.54 &7 &0.47  &6(d) \nl
4 &1286 &1.6749,4.6617 &ML  &-21.17$^{+0.19}_{-0.15}$ &3.53 &6 &0.73  &8(d) \nl
5 &763  &-0.2848,110.13 &BDO &-20.75$^{+0.16}_{-0.86}$ &3.31 &7 &0.85  &6(e) \nl
5 &763  &-0.2848,69.934 &ML  &-21.18$^{+0.19}_{-0.21}$ &4.92 &5 &0.42  &8(e) \nl
6 &523  &1.3754,10.240 &BDO &-21.15$^{+0.11}_{-0.92}$ &3.97 &7 &0.78  &6(f) \nl
6 &523  &1.3754,7.4759 &ML  &-21.46$^{+0.34}_{-0.29}$ &15.5 &6 &0.016  &8(f) \nl
7 &261  &2.3284,2.6809 &BDO &-21.45$^{+0.40}_{-0.53}$ &3.03 &7 &0.88  &6(j) \nl
8 &666  &1.5361,9.1237 &BDO &-21.25$^{+0.28}_{-0.45}$ &4.32 &7 &0.74  &6(g)\nl  
9 &620  &2.0242,4.6980 &BDO &-21.05$^{+0.20}_{-0.42}$ &3.49 &7 &0.83  &6(h) \nl 
10 &671  &0.5468,24.655 &BDO &-20.45$^{+0.37}_{-0.90}$ &3.05 &7 &0.88 &6(i) \nl 
\tablenotetext{\footnotesize{a}}{\footnotesize{(1) DB96 sample;
(2) all lines, Ly$\alpha$ QSO redshifts; (3) all lines, 
using narrow line redshifts where available, Ly$\alpha$ redshifts otherwise; 
(4) all lines, narrow line redshifts where available, Ly$\alpha$  
redshifts + 400 km s$^{-1}$ otherwise; (5) high z lines, QSO redshifts as
in case (4) ;  
(6) low z lines,  QSO redshifts as in case (4); 
(7) lines from QSOs with damped Ly$\alpha$ systems only, QSO redshifts as in
case (4);
(8) high luminosity QSOs, QSO redshifts as in case (4);
(9) low  luminosity QSOs, QSO redshifts as in case (4);
(10) weak lines only: $0.16 \AA < W < 0.32 \AA$, QSO redshifts as in case (4)}}
\tablenotetext{\footnotesize{b}}{\footnotesize{number of Ly$\alpha$ forest lines in sample}}
\tablenotetext{\footnotesize{c}}{\footnotesize{Equ.~\ref{eq:dndz} parameters 
$\gamma$ and ${\cal A}_{0}$ from
maximum likelihood fit to data; when the method listed is ML,
the normalization listed is equal to ${\cal A}_{0}(N_{lim}/N_{0})^{\beta-1}
(\beta-1)^{-1}$ (see text,Paper I)}}
\tablenotetext{\footnotesize{d}}{\footnotesize{BDO- Bajtlik, Duncan, \& Ostriker (1988), ML- maximum 
likelihood, see Kulkarni \& Fall (1993)}}
\tablenotetext{\footnotesize{e}}{\footnotesize{Best fit value of log[$(J(\nu_{0})$] in units of ergs s$^{-1}$
cm$^{-2}$ Hz$^{-1}$ sr$^{-1}$} }
\tablenotetext{\footnotesize{f}}{\footnotesize{$\chi^{2}$ of data vs. the ionization model used }}
\tablenotetext{\footnotesize{g}}{\footnotesize{number of points used to calculated $\chi^{2}$}}
\tablenotetext{\footnotesize{h}}{\footnotesize{$\chi^{2}$ probability for the ionization model used}}
\tablenotetext{\footnotesize{i}}{\footnotesize{Figure displaying number distribution per coevolving 
redshift interval, $d{\cal N}/dX_{\gamma}$}}
\enddata
\end{deluxetable}

\begin{deluxetable}{llcc}
\tablefontsize{\footnotesize}
\tablecolumns{4}
\tablewidth{28pc}
\tablecaption{Emission Line Redshifts for QSOs Included in Measurement of 
$J(\nu_{0})$\tablenotemark{a}  \label{table-z}  }
\tablehead{
\colhead{QSO}  &\colhead{z} &
\colhead{line\tablenotemark{b}}  &\colhead{Ref.\tablenotemark{c}}  }
\startdata
0000-263* &4.111 &Ly$\alpha$ &1  \nl
	  &4.116 &Mg II       &2 \nl
0001+087  &3.243 &Ly$\alpha$ &3  \nl
0002+051  &1.899 &Ly$\alpha$  &4  \nl
	  &1.899 &Mg II       &5   \nl 
0002-422  &2.763 &Ly$\alpha$  &6 \nl 
0006+020* &2.340 &Ly$\alpha$  &2  \nl
0014+813  &3.386 &Ly$\alpha$  &1 \nl
	  &3.379 &Mg II       &2  \nl
	  &3.404 &H$\beta$    &2  \nl
0027+014 &2.333  &Ly$\alpha$  &2  \nl 
	 &2.310  &H$\beta$    &7  \nl  
0029+073 &3.261  &Ly$\alpha$  &1 \nl
0037-018 &2.341  &Ly$\alpha$ &2  \nl
0049+007 &2.275  &Ly$\alpha$ &2   \nl
	 &2.279  &[OIII] $\lambda$5007  &8 \nl
0058+019 &1.959  &Ly$\alpha$ &9  \nl
	 &1.964  &Mg II      &5      \nl
0100+130 &2.690  &Ly$\alpha$ &6  \nl
0114-089 &3.194  &Ly$\alpha$ &9   \nl
	 &3.192  &Mg II      &2  \nl
0119-046* &1.951 &Ly$\alpha$ &1   \nl
	  &1.964 &Mg II  &5 \nl
0123+257* &2.358 &Ly$\alpha$ &10    \nl
	  &2.370 &[OIII] $\lambda\lambda$4959,5007 
          &8  \nl 
0142-100* &2.727  &Ly$\alpha$ &9  \nl
0150-202* &2.148  &Ly$\alpha$ &2   \nl
	  &2.149  &Mg II      &5  \nl
0153+744* &2.340  &Ly$\alpha$ &2   \nl 
	  &2.341  &[OIII] $\lambda$5007 &8 \nl
0226-038 &2.067  &Ly$\alpha$  &2  \nl 
	 &2.073  &Mg II &5 \nl
	 &2.073  &[OIII] $\lambda\lambda$4959,5007 &8 \nl
0237-233 &2.224  &Ly$\alpha$ &9      \nl
	 &2.200  &[OIII] $\lambda$5007 &7  \nl
0256-000 &3.374  &Ly$\alpha$   &1  \nl
0301-005 &3.228  &Ly$\alpha$   &1       \nl
0302-003*&3.286  &Ly$\alpha$   &1    \nl
0334-204 &3.131  &Ly$\alpha$   &1       \nl
0348+061 &2.057  &Ly$\alpha$ &2  \nl 
	 &2.056  &Mg II      &5  \nl
0400+258 &2.108  &Ly$\alpha$ &2   \nl
0421+019 &2.050  &Ly$\alpha$ &4  \nl
	 &2.056  &Mg II      &5  \nl
0424-131*&2.165  &Ly$\alpha$ &11   \nl
	 &2.166  &Mg II &5 \nl
	 &2.163  &H$\alpha$ &12  \nl
0453-423 &2.656  &Ly$\alpha$ &6  \nl 
0636+680 &3.167  &Ly$\alpha$ &1  \nl
	 &3.184  &Mg II &2  \nl
	 &3.187  &H$\beta$ &2  \nl
0731+653 &3.033  &Ly$\alpha$ &1  \nl
0747+610 &2.491  &Ly$\alpha$ &2   \nl
0831+128 &2.739  &Ly$\alpha$ &3 \nl
0836+710* &2.189  &Ly$\alpha$ &2 \nl 
	  &2.197  &Mg II  &2  \nl
	  &2.218 &[OIII] $\lambda$5007 &8 \nl
0837+109 &3.323 &Ly$\alpha$ &9  \nl
0848+155 &2.019 &Ly$\alpha$ &2   \nl 
	 &2.014 &Mg II &5 \nl
0848+163*&1.925 &Ly$\alpha$ &13  \nl
	 &1.922 &Mg II &5 \nl
0905+151 &3.173 &Ly$\alpha$ &3 \nl
0913+072 &2.785 &Ly$\alpha$ &9  \nl
0936+368 &2.025 &Ly$\alpha$ &2   \nl
0938+119 &3.192 &Ly$\alpha$ &3  \nl
0952+335 &2.504 &Ly$\alpha$ &2  \nl
0955+472 &2.482 &Ly$\alpha$ &2   \nl
0956+122 &3.033 &Ly$\alpha$ &2   \nl
	 &3.299 &Mg II &2 \nl
	 &3.314 &H$\beta$ &2 \nl
	 &3.308 &[OIII] $\lambda$5007 &2 \nl
1009+299 &2.633 &Ly$\alpha$ & 2 \nl
1017+280 &1.928 &Ly$\alpha$ &9  \nl
1033+137 &3.092 &Ly$\alpha$ &3  \nl
1115+080*&1.727 &Ly$\alpha$ &14  \nl
1159+124*&3.505 &Ly$\alpha$  &9     \nl
	 &3.508 &Mg II &2  \nl
	 &3.497 &H$\beta$ &2  \nl
	 &3.497 &[OIII] $\lambda$5007 &2  \nl
1206+119 &3.108 &Ly$\alpha$ &3  \nl
1207+399 &2.451 &Ly$\alpha$ &2  \nl 
	 &2.463 &H$\alpha$   &2   \nl
1208+101*&3.822 &Ly$\alpha$ &3  \nl
	 &3.833 &H$\beta$ &2  \nl
	 &3.802 &[OIII] $\lambda$5007 &2  \nl
1210+175*&2.564 &Ly$\alpha$  &2  \nl
1215+333 &2.606 &Ly$\alpha$  &14 \nl
1225-017 &2.831 &Ly$\alpha$ &15 \nl 
1225+317 &2.200 &Ly$\alpha$ &6  \nl
	 &2.226 &[OIII] $\lambda$5007 &8**  \nl
1231+294 &2.018 &Ly$\alpha$  &2   \nl
1247+267 &2.041 &Ly$\alpha$  &9  \nl
1315+472 &2.590 &Ly$\alpha$ &3  \nl
1323-107 &2.360 &Ly$\alpha$/C IV  &16  \nl
1329+412*&1.934 &Ly$\alpha$ &2 \nl 
1334-005 &2.842 &Ly$\beta$ &3  \nl
1337+285* &2.541 &Ly$\alpha$ &2  \nl
1346-036 &2.356 &Ly$\alpha$  &2   \nl 
	 &2.368 &Mg II &12  \nl 
	 &2.362 &[OIII] $\lambda$5007 &8  \nl
	 &2.367 &H$\alpha$ &12   \nl
1358+115*&2.589 &Ly$\alpha$ &2   \nl
1400+114 &3.177 &Ly$\alpha$ &3  \nl
1402+044 &3.208 &Ly$\alpha$ &17 \nl
1406+492 &2.161 &Ly$\alpha$ &2   \nl
1408+009 &2.262 &Ly$\alpha$ &2   \nl 
	 &2.260 &[OIII] $\lambda$5007 &2  \nl
	 &2.265 &H$\alpha$ &2  \nl
1410+096 &3.313 &Ly$\alpha$ &3  \nl
1421+330 &1.903 &Ly$\alpha$ &2   \nl 
	 &1.906 &Mg II &5  \nl
1422+231*&3.624 &Ly$\alpha$ &2   \nl
	 &3.630 &H$\beta$ &2   \nl
	 &3.623 &[OIII] $\lambda$5007 &2   \nl
1435+638 &2.063 &Ly$\alpha$ &2   \nl
	 &2.061 &Mg II &5 \nl 
	 &2.066 &[OIII] $\lambda\lambda$4959,5007 &8  \nl
	 &2.065 &H$\alpha$ &2  \nl 
1442+101 &3.560 &Ly$\alpha$ &17  \nl 
1451+123 &3.251 &Ly$\alpha$ &3  \nl
1511+091*&2.877 &C IV       &9    \nl
1512+132 &3.120 &Ly$\alpha$ &3 \nl
1548+092 &2.759 &Ly$\alpha$ &9  \nl
1601+182*&3.227 &Ly$\alpha$ &3   \nl
1602+178*&2.989 &Ly$\alpha$ &3   \nl
1603+383*&2.510 &Ly$\alpha$ &18 \nl
1604+290 &1.962 &Ly$\alpha$ &2   \nl
1607+183 &3.120 &Ly$\alpha$ &17  \nl
1614+051 &3.216 &Ly$\alpha$ &17  \nl
	 &3.214 &[OIII] $\lambda$5007 &19  \nl
1623+269*&2.526 &Ly$\alpha$ &3  \nl
1700+642 &2.744 &Ly$\alpha$ &15  \nl
1715+535 &1.935  &Ly$\alpha$ &2   \nl
	 &1.932 &Mg II &5  \nl
1738+350 &3.239 &Ly$\alpha$ &3 \nl
1946+770 &3.020 &Ly$\alpha$ &15 \nl
2126-158 &3.280 &Ly$\alpha$ &6  \nl
	 &3.284 &Mg II &2  \nl
	 &3.298 &H$\beta$ &2  \nl
         &3.292 &[OIII] $\lambda$5007  &2  \nl
2134+004 &1.941 &Ly$\alpha$ &2   \nl 
2233+131 &3.301 &Ly$\alpha$ &1  \nl
2233+136 &3.207 &Ly$\alpha$ &1  \nl
2251+244*&2.335 &Ly$\alpha$ &17  \nl 
	 &2.359 &[OIII] $\lambda$5007 &8  \nl
2254+024 &2.089 &Ly$\alpha$ &2  \nl 
	 &2.090 &Mg II &5 \nl
2310+385 &2.179 &Ly$\alpha$ &2   \nl 
	 &2.181 &[OIII] $\lambda\lambda$4959,5007 &8  \nl
2311-036 &3.041 &Ly$\alpha$ &1  \nl
2320+079 &2.088 &Ly$\alpha$ &2   \nl
2329-020 &1.896 &Ly$\alpha$ &2  \nl
\enddata
\tablenotetext{\footnotesize{a}}{{Objects with both [OIII] and Ly$\alpha$, Mg II, or Balmer
redshifts were used to construct histograms in Figure~\ref{fig:zhist}; 
Objects
marked with an asterisk are excluded from the proximity effect analysis on
the basis of associated absorption or gravitational lensing}}
\tablenotetext{\footnotesize{b}}{\footnotesize{Emission lines used to measure redshift.}}
\tablenotetext{\footnotesize{c}}{\footnotesize{REFERENCES: 
(1) Sargent et al. 1989; 
(2) this paper;
(3) B94;
(4) Young et al. 1982a;
(5) Steidel \& Sargent 1991;
(6) Sargent et al. 1980;
(7) Baker et al. 1994;  
(8)M$^{\rm c}$Intosh et al. 1999a (**1225+317 measurement quoted as uncertain
due to weak [O III] emission and low S/N);
(9) Sargent et al. 1988;
(10) Schmidt \& Olsen 1968;
(11) Burbidge 1970;
(12) Espey et al. 1989;
(13) Young et. al 1982b;
(14) Wills \& Wills 1979;
(15) DB96;
(16) Kunth et al. 1981;
(17) Barthel et al. 1990;
(18) Hamburg/CfA Bright Quasar Survey (unpublished);
(19) Bremer \& Johnstone 1995} }
\end{deluxetable}

\begin{deluxetable}{lccc}
\tablefontsize{\footnotesize}
\tablewidth{35pc}
\tablecaption{Ionization Rates \label{table-hm}} 
\tablehead{
\colhead{$A,B,z_{c},S$}   &\colhead{Ref.} &
\colhead{$\chi^{2}$}  &\colhead{Q$_{\chi^{2}}$} \\
\colhead{}  & \colhead{(a)} &\colhead{(b)} &\colhead{(c)} }
\startdata
6.7e-13 s$^{-1}$,0.43,2.30,1.95 &1 &11.2 &0.12 \nl
5.6e-13 s$^{-1}$,0.60,2.22,1.90 &2 &11.8 &0.10 \nl
1.2e-12 s$^{-1}$,0.58,2.77,2.38 &2 &7.15 &0.41 \nl
\tablenotetext{\footnotesize{a}}{\footnotesize{(1)Haardt \& Madau (1996); 
(2)Fardal et al. (1998)}}
\tablenotetext{\footnotesize{b}}{\footnotesize{$\chi^{2}$ of data vs. the BDO ionization model }}
\tablenotetext{\footnotesize{c}}{\footnotesize{$\chi^{2}$ probability for the  BDO ionization model}}
\enddata
\end{deluxetable}

\begin{deluxetable}{ccccc}
\tablefontsize{\footnotesize}
\tablecolumns{5}
\tablewidth{40pc}
\tablecaption{Simulation Results \label{table-sim}} 
\tablehead{
\colhead{Input log[$(J(\nu_{0})$]} &\colhead{$\gamma$,${\cal A}_{0}$}   &
\colhead{log[$(J(\nu_{0})$] recovered} &\colhead{$\chi^{2}$} &
\colhead{Q$_{\chi^{2}}$} \\ 
\colhead{(a)}  & \colhead{(b)} & \colhead{(c)} &\colhead{(d)} &\colhead{(e)} }
\startdata
-23.0 &1.5722,11.043  &-22.75$^{+0.28}_{-0.19}$ &11.2 &0.12   \nl
-22.0 &1.6869,8.8367  &-21.80$^{+0.40}_{-0.28}$ &11.3 &0.12   \nl
-21.3 &2.6267,2.6960  &-21.00$^{+0.28}_{-0.60}$ &2.68 &0.91   \nl
-20.0 &2.2511,3.8084  &-19.50$^{+1.86}_{-0.84}$ &3.90 &0.79   \nl
-19.0 &2.0302,5.2704  &-18.50$^{+0.66}_{-1.56}$ &5.28 &0.62   \nl
\tablenotetext{\footnotesize{a}}{\footnotesize{value of log[$(J(\nu_{0})$] used for modifying absorber column
densities according to Equations~\ref{eq:column} and~\ref{eq:omega}} }
\tablenotetext{\footnotesize{b}}{\footnotesize{Equ.~\ref{eq:dndz} parameters $\gamma$ and ${\cal A}_{0}$
from maximum likelihood fit to data}}
\tablenotetext{\footnotesize{c}}{\footnotesize{value of log[$(J(\nu_{0})$] from simulated spectra using the
standard BDO technique}}
\tablenotetext{\footnotesize{d}}{\footnotesize{$\chi^{2}$ of data vs. the BDO ionization model }}
\tablenotetext{\footnotesize{e}}{\footnotesize{$\chi^{2}$ probability for the BDO ionization model}}
\enddata
\end{deluxetable}

\begin{deluxetable}{lccc}
\tablefontsize{\footnotesize}
\tablecolumns{4}
\tablewidth{40pc}
\tablecaption{Literature Proximity Effect Measurements of $J(\nu_{0})$ 
\label{table-lit}} 
\tablehead{
\colhead{log[$(J(\nu_{0})$]} &\colhead{z}   &
\colhead{${\cal N}_{QSOs}$} 
&\colhead{Ref.}\tablenotemark{a}}
\startdata
-23.2$^{+0.8}_{-0.6}$ &0.16-0.99  &13   &1  \nl
-20.5$^{+\infty}_{-1.3}$ &1.8-2.3 &3   &2\tablenotemark{b} \nl
-21.1$\pm$0.6     &2.0 &1    &2 \nl
-21.15              &3.2 &1     &3 \nl
-21.3               &3.6 &1     &4 \nl
-21.0$\pm$0.5       &1.7-3.8 &38    &5 \nl
-21.0$\pm$0.5       &1.7-3.8 &19    &6 \nl
-20.5               &1.6-4.1 &49    &7 \nl
-21.3$^{+0.08}_{-0.09}$ &1.7-4.1  &10    &8 \nl
-21.1$^{+0.15}_{-0.27}$ &1.7-4.1  &74   &9  \nl 
-21.0$^{+0.17}_{-0.15}$ &2.0-4.5  &11   &10  \nl
-22.0 - -21.5       &4.5 &1   &11 \nl
\tablenotetext{\footnotesize{a}}{\footnotesize{
(1) Kulkarni \& Fall 1993;
(2) Fern\'{a}ndez-Soto et al. 1995;
(3) Giallongo et al. 1993;
(4) Cristiani et al. 1995;
(5) BDO:
(6) LWT;
(7) B94;
(8) Giallongo et al. 1996;
(9) this paper;
(10) Cooke et al. 1997;
(11) Williger et al. (1994)}}
\tablenotetext{\footnotesize{b}}{\footnotesize{measured from the proximity effect to due a foreground QSO;
not able to set an upper limit}}
\enddata
\end{deluxetable}

\end{document}